\documentclass{article}%[draft]
\usepackage{PRIMEarxiv}
\usepackage[style=numeric, backend=biber]{biblatex}
\usepackage{amsmath,amssymb,amsfonts}
\usepackage[noend]{algorithmic}
\usepackage{graphicx}
\usepackage{textcomp}
\usepackage{xcolor}
\def\BibTeX{{\rm B\kern-.05em{\sc i\kern-.025em b}\kern-.08em
    T\kern-.1667em\lower.7ex\hbox{E}\kern-.125emX}}

\usepackage{url}

\usepackage{booktabs}       % professional-quality tables
\usepackage[flushleft]{threeparttable}
\usepackage{amsfonts}       % blackboard math symbols
\usepackage{nicefrac}       % compact symbols for 1/2, etc.
\usepackage{microtype}      % microtypography
\usepackage{lipsum}
\usepackage{fancyhdr}       % header
\usepackage{physics}
\usepackage{mathtools}
\usepackage{xcolor}
\usepackage{soul}
\usepackage{multirow}
\usepackage{textcomp}

\usepackage{enumitem}

\usepackage[ruled, linesnumbered, vlined]{algorithm2e}

\usepackage{mathabx}
\allowdisplaybreaks

\usepackage{hyperref}       % hyperlinks
\hypersetup{
    colorlinks=true,
    linkcolor=blue,
    filecolor=magenta,      
    urlcolor=cyan,
}

\usepackage{soul}

\usepackage{lipsum}

\newcommand\blfootnote[1]{%
  \begingroup
  \renewcommand\thefootnote{}\footnote{#1}%
  \addtocounter{footnote}{-1}%
  \endgroup
}

\usepackage{lineno}
\usepackage{setspace}
\begin{document}

\title{{Trajectory-Integrated Accessibility Analysis of\\Public Electric Vehicle Charging Stations }
% {\footnotesize \textsuperscript{*}Note: Sub-titles are not captured in Xplore and
% should not be used}

}

% \makeatletter
% \newcommand{\linebreakand}{%
%   \end{@IEEEauthorhalign}
%   \hfill\mbox{}\par
%   \mbox{}\hfill\begin{@IEEEauthorhalign}
% }
% \makeatother

% \author{
% \emph{anonymous authors}
% }

\author{
Yi Ju\\
% \textit{Department of Civil and Environmental Engineering} \\
UC Berkeley\\
% Berkeley, CA, USA \\
% \texttt{juy16thu@berkeley.edu}\\
\And
Jiaman Wu \\
% Department of Civil and Environmental Engineering \\
UC Berkeley \\
% Berkeley, CA, USA\\
% \texttt{jmwu@berkeley.edu} \\
\And
Zhihan Su \\
% Department of Building Science and Technology \\
Tsinghua University \\
% Beijing, China\\
% \texttt{?} \\
\And
Lunlong Li \\
% Department of Building Science and Technology \\
HKUST \\
% Singapore\\
% \texttt{lunlong.li@connect.ust.hk} \\
\AND
Jinhua Zhao \\
% Department of Civil and Environmental Engineering \\
M.I.T. \\
% Cambridge, MA, USA\\
% \texttt{jinhua@mit.edu} \\
\And
Marta C. Gonz\'{a}lez \\
% Department of Civil and Environmental Engineering \\
UC Berkeley \\
% Berkeley, CA, USA\\
% \texttt{martag@berkeley.edu} \\
\And
Scott Moura \textsuperscript{\textdagger} \\
% Department of Civil and Environmental Engineering \\
UC Berkeley
% Berkeley, CA, USA\\
% \texttt{smoura@berkeley.edu} \\
}

\maketitle

\blfootnote{
Contact: \texttt{juy16thu@berkeley.edu} (Y. Ju), \texttt{smoura@berkeley.edu} (S. Moura). \textdagger: corresponding author.
\\Published at \textit{Sustainable Cities and Society}: \href{https://doi.org/10.1016/j.scs.2026.107491}{doi:10.1016/j.scs.2026.107491}.
This version updated on: \emph{May 18, 2026}.
}

\begin{abstract}
Electric vehicle (EV) charging infrastructure is crucial for advancing EV adoption, managing charging loads, and ensuring equitable transportation electrification. However, there remains a notable gap in comprehensive accessibility metrics that integrate the mobility of the users. This study introduces a novel accessibility metric, termed Trajectory-Integrated Public EVCS Accessibility (TI-acs), and uses it to assess public electric vehicle charging station (EVCS) accessibility for approximately 6 million residents in the San Francisco Bay Area based on detailed individual trajectory data in one week. Unlike conventional home-based metrics, TI-acs incorporates the accessibility of EVCS along individuals' travel trajectories, bringing insights on more public charging contexts, including public charging near workplaces and charging during grid off-peak periods. 

As of June 2024, given the current public EVCS network, Bay Area residents have, on average, 7.5 hours and 5.2 hours of access per day during which their stay locations are within 1 km (i.e. 10-12 min walking) of a public L2 and DCFC charging port, respectively. Over the past decade, TI-acs has steadily increased from the rapid expansion of the EV market and charging infrastructure. However, spatial disparities remain significant, as reflected in Gini indices of 0.38 (L2) and 0.44 (DCFC) across census tracts. Additionally, our analysis reveals racial disparities in TI-acs, driven not only by variations in charging infrastructure near residential areas but also by differences in their mobility patterns.

\end{abstract}


% \section{Introduction}\label{sec:intro}

\vspace{20pt}
\section{Introduction}
\subsection{Background}
The transportation sector is undergoing a transformative shift toward electrification and decarbonization. Electric vehicle (EV) sales and inventories are rapidly growing in major global markets, including the United States \cite{bibra2022electric}. In California alone, over 1.2 million battery electric vehicles (BEVs) had been sold by 2023 \cite{cec_ev_sales_2024}, a trend expected to accelerate under the state’s Advanced Clean Cars II regulation, which mandates 100\% zero-emission vehicle (ZEV) sales by 2035 \cite{ACCII_2024}. A critical enabler of this transition is the availability of accessible, efficient, and reliable charging infrastructure.

Empirical studies in California have demonstrated that proximity to EV charging stations (EVCS) positively affects local housing markets \cite{liang2023effects} and nearby businesses \cite{zheng2024effects}, by increasing traffic flow, attracting visitors, and improving air quality \cite{yu2023california} (through greater EV adoption \cite{ledna2022support}). 
Yet, \emph{charging anxiety} remains a significant barrier to broader adoption. As of 2023, the U.S. had approximately a ratio of 26 light-duty BEVs per public EV charger \cite{iea2024evoutlook}, and nationwide surveys identified ``public charger availability'' and ``public charger reliability'' as the top concerns among EV drivers \cite{plug2024survey}. Moreover, disparities in charger accessibility across racial and income groups have been documented \cite{hsu2021public, lou2024income}, though such studies often rely on coarse metrics, such as the number of charging ports within predefined geographic units.

This paper advances the study of public EVCS accessibility by introducing novel, \emph{mobility-informed} accessibility metrics grounded in human travel behaviors. 
Our approach aims to promote recent efforts that translate large-scale movement data into urban sustainability indicators \cite{ozturk2025mesoscopic}, and 
provide deeper insights into the design and deployment of charging infrastructure, with implications for sustainability \cite{zhang2024sustainable, powell2022charging}, grid stability \cite{navidi2023coordinating, elmallah2022can}, and social equity \cite{hsu2021public, brockway2021inequitable}.

\subsection{Related works}\label{sec:lit_rev}

The pivotal role of EV charging infrastructure in supporting personal vehicle electrification has motivated a growing body of research on its accessibility. Numerous case studies have examined this issue across leading EV markets, including the United States \cite{lou2024income, carlton2024electric, gazmeh2024understanding, bashar2026evaluating}, China \cite{peng2024analytical, li2022spatial}, and Europe \cite{falchetta2021electric}, as well as at more localized scales such as California \cite{hsu2021public, roy2022examining}, Texas \cite{jiao2024toward}, Washington \cite{esmaili2024assessing}, Montreal \cite{farnood2026proximity}, and New York City \cite{khan2022inequitable}. 
While these studies are unified in their goal of addressing charging access as a critical concern, they differ considerably in their quantitative frameworks for defining and evaluating accessibility, reflecting diverse methodological choices and regional contexts.
A central question in defining accessibility quantitatively lies in determining how individuals are matched with EV charging station (EVCS) infrastructure — that is, identifying which EVCS locations are considered accessible to a given person or population group.

The most common approaches are based on individuals’ home locations, often acquired from census data on residential populations. Several studies adopt a \emph{count-based} method, using the number of EVCS within specific geographic units, such as census tracts or zip codes, as a proxy for charging accessibility of local residents. For example, Khan \textit{et al.} \cite{khan2022inequitable} quantified the number of EVCS within each zip code in New York City and examined how these counts correlate with the demographic characteristics of the corresponding resident populations. 
Hsu \textit{et al.} \cite{hsu2021public} conducted a case study in California in which they defined accessibility as the existence of at least one public charger with the boundary of census block group. Then they compared the probabilities of having public charger access among different demographic groups.
A key limitation of the count-based approach lies in its sensitivity to the spatial scale of analysis. When the geographic unit is too large (e.g., a county), EVCS may be located far from many residents within the same unit, overstating actual accessibility. In contrast, when the unit is too small (e.g., a census block group), most areas contain no public EVCS at all, leading to sparse or misleading measures. In both cases, such metrics fail to accurately reflect individuals’ actual access to charging infrastructure.

\emph{Distance-based} measures offer an alternative to count-based approaches by assessing accessibility in terms of the travel distance or travel time between individuals and the nearest EVCS. For example, Carlton \textit{et al.} \cite{carlton2024electric} calculated travel time to the nearest EVCS for each pixel across the contiguous United States and aggregated the results to the census tract level. Similarly, Lou \textit{et al.} \cite{lou2024income} leveraged a disaggregated dataset of 121 million U.S. household locations to compute the shortest distance to up to five nearby EVCSs for each household. 

While distance-based methods typically consider the nearest one or few charging stations to their homes, in reality, individuals may access a broader set of EVCSs, with the general tendency that the likelihood of use decreases with distance. Building on this insight, \emph{gravity-based} models offer a more nuanced approach by capturing how accessibility decays over space. These models compute a weighted sum over multiple EVCSs, assigning lower influence to stations that are farther away. Esmaili \textit{et al.} \cite{esmaili2024assessing} applied a gravity model to evaluate EVCS accessibility at the census tract level in King County, Washington, incorporating a rivalry factor to represent competition for charging resources. Similarly, Peng \textit{et al.} \cite{peng2024analytical} employed the two-step floating catchment area (2SFCA) model to assess public charger accessibility in Hong Kong. This model introduces a more symmetric view: charging stations compete for multiple EVs (supply-side competition), and EVs can access multiple stations (demand-side flexibility), with both interactions weighted by distance decay. A related concept is the Public EV Charging Opportunity (PECO) score proposed by Zhang \textit{et al.} \cite{zhang2025equity}, which quantifies the equity of community-level EVCS service provision in the Sacramento region using a similar distance-weighted supply-demand framework.  

Although the aforementioned studies differ in their choice of metrics and data sources, they share a common limitation: accessibility is typically restricted to public chargers located near individual homes, whether within the same geographic unit or among the nearest few stations. This home-centric perspective overlooks the fact that many charging events occur while individuals are away from home, such as at workplaces, retail centers, or other public destinations. Moreover, public EVCSs located in commercial and industrial areas often exhibit higher utilization rates, improving cost-revenue efficiency \cite{hamim2025real}. These locations also play a strategic role in grid decarbonization by aligning charging demand with daytime solar generation. Such heterogeneity motivates accessibility metrics that reflect where people actually spend time, not only where they live.

Incorporating charging opportunities around various locations of activity, beyond the home, can provide a more comprehensive understanding of the accessibility of the charger. However, such analysis has traditionally been constrained by the limited availability of detailed \emph{human mobility data}. Among the few studies addressing this gap, 
Gazmeh \textit{et al.} \cite{gazmeh2024understanding} incorporated Point-of-Interest (POI) data from SafeGraph to develop opportunity-centric measures of accessibility. They introduced two metrics: activity-induced charging accessibility (AICA), which quantifies access to nearby EVCS when visiting a specific category of POIs, and charging-induced activity accessibility (CIAA), which measures access to POIs when charging at EVCS. Their analysis encompassed all census block groups (CBGs) within the 20 largest US metropolitan areas, covering over 5 million POIs across nine major categories. The study offered valuable insights into the interaction between charging behavior and daily activities. However, because individuals often visit POIs outside the CBGs where they reside, correlating these accessibility measures with local demographic characteristics may produce misleading interpretations.
Kontou \textit{et al.} \cite{kontou2019understanding} analyzed two GPS datasets that contain longitudinal driving trajectories from 275 and 1,061 households, respectively. They introduced two novel accessibility metrics, namely stop-based and daily trip chain–based charging opportunities, which account for the locations and timing of trip stops. However, the utility of the study is restricted not only by its limited sample size, but also by the absence of real-world EVCS data; instead, chargers were artificially placed at locations with high trip-stop density as a heuristic for infrastructure planning.

Recent studies have begun to move EVCS accessibility beyond purely home-based proximity. Qian \textit{et al.} \cite{qian2025accessibility} propose a visit-based charging accessibility framework that links charging opportunities to routine activity participation, and aggregate statistics across POI categories. Using large-scale PCS/POI and mobility-pattern datasets across multiple U.S. metropolitan areas, they show that although PCS may appear spatially close to lower-income communities, functional alignment between routine activities and PCS locations can still be worse for disadvantaged groups
In parallel, Mehditabrizi \textit{et al.} \cite{mehditabrizi2025route} extend accessibility measurement to consider both home proximity and en-route opportunities, emphasizing that equity conclusions can change once daily travel patterns are incorporated. More specifically on non-home destinations, Cai \textit{et al.} \cite{cai2026revisiting} evaluate accessibility under a home–work dual-scenario and report that including workplace opportunities can substantially change overall accessibility and inequality metrics. Wang \textit{et al.} \cite{wang2026leveraging} further leverage commuting and workplace charging to develop an equity-aware framework (and planning implications) centered on commuting-linked opportunities. Complementing these measurement efforts, recent planning studies also explicitly balance efficiency and equity objectives in EVCS deployment \cite{liu2025planning}.

Despite these advances, fully integrating mobility and infrastructure in a time-geographic manner remains underexplored. Existing mobility-aware approaches often rely on aggregated origin–destination/CBG-to-POI flows or scenario-based proxies rather than computing accessibility at the individual trajectory level, which limits flexible distributional analyses (e.g., low-percentile access relevant for early adopters) and complicates downstream behavioral extensions such as SOC-aware charging feasibility. In addition, many formulations represent charger supply via simplified presence/buffer logic around destinations and do not directly couple accessibility to time budgets across all daily activity stops (including home/work and other purposes) or to temporal availability considerations that matter for policy and grid-friendly charging (e.g., peak vs off-peak opportunity).

To address the lack of large-scale real-world individual trajectory data, several studies have employed \emph{synthetic} human mobility datasets. Xu \textit{et al.} \cite{xu2018planning} and Wu \textit{et al.} \cite{wu2024planning} examined the potential benefits of implementing customized charging schedules to support grid load management, using simulated one-week mobility data of San Francisco Bay Area residents. 
Both studies leveraged mobility data to estimate energy demand and identify feasible charging opportunities. Xu \textit{et al.} assumed drivers had access to charging infrastructure at home and/or at work. Wu \textit{et al.} extended this framework by incorporating public charging stations and adopting a more advanced behavioral model \cite{powell2022charging} to reflect heterogeneous charging preferences. Rather than making minor adjustments to the start or end times of charging sessions as in \cite{xu2018planning}, Wu \textit{et al.} proposed shifting entire sessions away from peak grid hours to off-peak periods, while still meeting drivers' state-of-charge needs and respecting their travel schedules. Despite these advancements, both studies either omitted public chargers entirely or modeled them in a highly aggregated manner with only total public charging capacity within each ZIP code, which limits the spatial resolution of accessibility modeling.

We summarize the key \textbf{research gaps} in the EVCS accessibility literature as follows:
\begin{itemize}
    \item Home-centric accessibility dominates: Most EVCS accessibility studies still focus on public chargers near individuals’ residences, which systematically overlooks charging opportunities encountered near other major activity destinations (e.g., workplaces, shopping, services) along daily travel patterns.
    \item Mobility-aware studies often simplify infrastructure supply: Even when travel schedules or activity patterns are incorporated, the charging network is frequently represented in an overly coarse way (e.g., simplified spatial coverage or homogeneous supply), rather than using the observed, highly heterogeneous spatial distribution (and capacity) of real-world EVCS.
    \item Mobility, infrastructure, and equity remains under-integrated: As a result, the joint effects of where chargers are and how people move and dwell are not fully integrated into accessibility analysis—limiting actionable insights into socio-economic and racial disparities in public-charging access, and obscuring how unequal mobility patterns and uneven infrastructure placement jointly shape who benefits from public charging.
\end{itemize}

\subsection{Paper overview \& main contributions}

\begin{figure}[h]  % 'h' means here, but LaTeX may move it depending on space
    \centering
    \includegraphics[width=\textwidth]{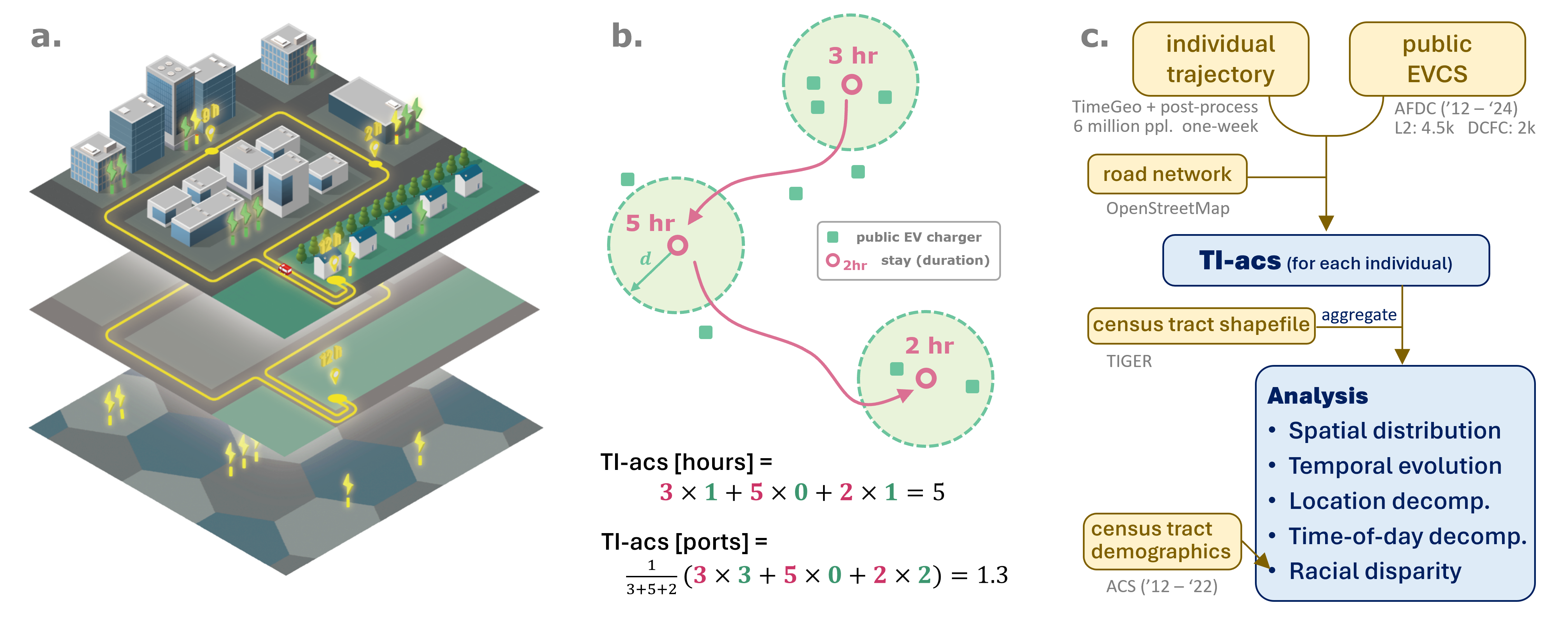}  % Adjust width or use scale instead
    \caption{
    \textbf{Graphic overview of the work.}
    {\small\qquad
    \textbf{a} Our research (top) integrates two data layers: one (bottom) is the geographical distribution of public charging resources, the other (middle) is week-long individual trajectories.\qquad
    \textbf{b} An illustrative example of TI-acs calculation. \qquad
    \textbf{c} Workflow of processing data, calculating TI-acs, and analyzing aggregated TI-acs. Yellow boxes are data types, with gray text below them annotating specific data sources.}
    }
    \label{fig:overview}  % Label for referencing the figure in the text
\end{figure}

In this paper, we introduce a novel class of accessibility metrics, termed \textbf{Trajectory-Integrated Public EVCS Accessibility (TI-acs)}, designed to provide a more comprehensive understanding of a driver's access to charging infrastructure, and especially public charging infrastructure. TI-acs quantifies public EVCS accessibility by incorporating individual mobility patterns, accounting for charging opportunities near homes, workplaces, and other locations where they spend time (e.g., entertainment venues, grocery stores).

As illustrated in Fig.\ref{fig:overview}.\textbf{a}, the core idea behind TI-acs is to evaluate public charging accessibility \emph{along an individual’s trajectory} over a given period. By identifying public chargers within a specified distance of each stay location and weighting them by duration (``integral’’), TI-acs effectively integrates charging opportunities across multiple daily activities. In this study, we focus on a key variant, \textbf{TI-acs [hours]}, also referred to as \emph{accessible hours}, which represents the total daily duration during which at least one public charger is available within a given proximity.

To the best of our knowledge, this framework is among the \emph{first of its kind}
to integrate individual mobility trajectories with detailed geographic data on public EVCS locations, offering deeper insights than either dataset alone. Our analysis covers approximately 6 million residents and more than 6,000 public charging ports in 1,500 census tracts in the San Francisco Bay Area. TI-acs is computed for each individual using week-long trajectory data mapped onto a detailed road network with public chargers placed at their precise locations.

To further dissect TI-acs, we analyze its \emph{composition} from multiple perspectives. One breakdown considers stay types, categorizing time spent at home, workplaces, and other locations. Another breakdown examines time-of-day variations in relation to electric utility tariff periods, including peak, off-peak, and super off-peak hours. This analysis explores the potential for shifting charging demand to reduce peak demand, without compromising mobility constraints.

From a \emph{temporal} perspective, we calculate TI-acs for annual snapshots of EVCS distribution from 2012 to 2024, capturing its evolution over the past decade, which complements recent efforts to characterize growth patterns and drivers of charging infrastructure expansion \cite{hu2025growth}. From a \emph{spatial} perspective, we aggregate individual TI-acs at the census tract level. Specifically, we assess its spatial distribution using the Gini index, and examine racial disparities by linking it to census tract demographics.

\textbf{Main contributions} of this paper are: \begin{enumerate}
    \item Propose a trajectory-integrated EVCS accessibility metric that quantifies public-charging opportunities along individuals’ daily activity–travel patterns, moving beyond residence-based proximity measures.
    \item Conduct a city-scale empirical assessment that combines high-resolution mobility trajectories with the observed spatial distribution (and capacity) of public charging infrastructure to characterize where, when, and for whom public charging is practically accessible.
    \item Reveal socio-economic and racial disparities in public-charging accessibility that arise from the coupled effects of uneven infrastructure placement and heterogeneous mobility/dwell patterns, providing equity-relevant evidence to inform planning and policy.
    \item Trace the evolution of trajectory-integrated accessibility over time using a longitudinal analysis, connecting infrastructure expansion to distributional changes in access and highlighting persistent or shifting inequities.
\end{enumerate}

\section{Methods}

\subsection{Trajectory-integrated accessibility}

\newcommand{\ts}{t^\text{s}}
\newcommand{\te}{t^\text{e}}
\newcommand{\traj}{\tau}
\newcommand{\Ns}{\mathcal{N}}
\newcommand{\Ss}{\mathcal{S}}
\newcommand{\Ts}{\mathcal{T}}
\newcommand{\Is}{\mathcal{I}}

Formally, the trajectory of an individual $i$ over a horizon $[0,T]$ (e.g., one week) can be formalized as a mapping $f_i: [0, T] \mapsto \Ss \cup \{\circ\}$, where $\Ss$ is the set of all locations, and $\circ$ means the vehicle is in movement. Denote $\Ns(s; d)$ as the set of available public charging ports whose distances to location $s$ are within $d$, and $\abs{\Ns(s;d)}$ is the cardinality of the set, i.e., the number of charging ports that meet the restriction. We only count destination charging opportunities, i.e., $\Ns(\circ; d)=\emptyset,\, \forall d\ge 0$. Let $\mathbf{1}\{x\}$ be the indicator function such that $\mathbf{1}\{x\}=1$ if $x$ is true, otherwise $0$.

Two accessibility metrics: \textbf{TI-acs [ports]} and \textbf{TI-acs [hours]}, are defined as:
\begin{subequations}\label{eq:TI-defn}
\begin{align}
    \text{[ports]}\qquad
        & \text{TI-acs}^{(d)}_i \coloneqq \frac{1}{T}\, \int_{t=0}^T\, \abs{\Ns\left(f_i(t); d\right)}\, {\rm d} t\\
    \text{[hours]}\qquad
        & \text{TI-acs}^{(d)}_i \coloneqq \int_{t=0}^T\, \mathbf{1}\{\abs{\Ns\left(f_i(t); d\right)}\ge 1\}\, {\rm d} t
\end{align}
\end{subequations}

In words, TI-acs [ports] reflects the stay-duration-weighted average \emph{number} of public charging ports accessible within walking distance $d$ from stay locations, while TI-acs [hours] quantifies the total \emph{duration} when \emph{at least} one public charger is within distance $d$.
TI-acs is evaluated separately for L2 (slow, $\sim$ 6 kW) and DCFC (fast, $\ge$ 50 kW), with 1 km as the primary distance threshold, which roughly corresponds to a 12–15 minute walk.

\paragraph{A toy example} In Fig.\ref{fig:overview}.\textbf{b}, pink circles connected with pink curves represent stays along a trajectory; green squares are the locations of public EV chargers. Accessible chargers are those within the light green circles centered at each stay, with a diameter $d$. In this example, the EV driver stays at their first stop for 3 hours, where there are 3 chargers nearby. They move to the second and third stops subsequently, 5 hours at the second stop, with no chargers within range $d$, and 2 hours at the third stop, with 2 chargers within range $d$.
Following the definition\footnote{For simplicity, we ignore the travel times, which have an impact on TI-acs [ports].}, 
\textbf{TI-acs [hours]} = $ 3 \times \mathbf{1}\{\texttt{True}\} + 5 \times \mathbf{1}\{\texttt{False}\} + 2 \times \mathbf{1}\{\texttt{True}\} = 5$ hours, while 
\textbf{TI-acs [ports]} = $\frac{1}{3+5+2}\, (3\times 3 + 5 \times 0 + 2\times 2) = 1.3$ ports.

\paragraph{Computational implementation.}
In our dataset, the trajectory of individual $i$ is represented by the data structure $\traj_i = \{(\ts_k, \te_k, s_k)\}_{k=1}^{K_i}$, 
where $\ts_k, \te_k, s_k$ is the start time, end time, and location of stay $k$, respectively.
In practice, the metric is calculated by the following equivalent definition:
\begin{subequations}\label{eq:TI-defn2}
\begin{align}
    \text{[ports]}\qquad
        & \text{TI-acs}^{(d)}_i \coloneqq \frac{1}{T}\, \sum\nolimits_{k \in \traj_i}\, \abs{\Ns(s_k; d)}\, \lambda_k\\
    \text{[hours]}\qquad
        & \text{TI-acs}^{(d)}_i \coloneqq \sum\nolimits_{k \in \traj_i}\, \mathbf{1}\{\abs{\Ns\left(s_k; d\right)}\ge 1\}\, \lambda_k
\end{align}
\end{subequations}

 where $\lambda_k = \te_k - \ts_k$ is the duration of each stay.
Here we abuse the notation $k \in \traj_i$ a bit for ``index $k$ of stays in trajectory $\traj_i$''.

\paragraph{TI-acs defined for a segment.}

When the accessibility of a particular segment is of interest, for instance, accessibility near one's workplace, or accessibility during off-peak hours of the grid, it is convenient to adapt the definition in eq.\eqref{eq:TI-defn} as:

\begin{subequations}\label{eq:TI-defn-seg}
\begin{align}
    \text{[ports]}\qquad
        & \text{TI-acs}^{(d)}_i[\Ts] \coloneqq \frac{1}{\abs{\Ts}}\, \int_{t\in\Ts}\, \abs{\Ns\left(f_i(t); d\right)}\, {\rm d} t\\
    \text{[hours]}\qquad
        & \text{TI-acs}^{(d)}_i[\Ts] \coloneqq \int_{t\in\Ts}\, \mathbf{1}\{\abs{\Ns\left(f_i(t); d\right)}\ge 1\}\, {\rm d} t
\end{align}
\end{subequations}

and each metric can be viewed as TI-acs$[\Ts]$ with a specific $\Ts \subseteq [0,T]$. 
For instance, $\Ts = [0,T]$ defines the standard TI-acs that integrates all segments, $\Ts = \{t\in[0,T]: f_i(t)=\text{workplace}_i\}$ defines accessibility near one's workplace, and $\Ts = \{t\in[0,T]: t~\text{is during off-peak hours}\}$ defines accessibility during off-peak hours of the grid, etc.

Although we do not use this formula for calculation in practice, it provides a unified perspective on defining various accessibility metrics, which deepens our understanding of the composition of public charger accessibility across one's daily schedule.

\paragraph{Aggregation methods.}
TI-acs is calculated for each individual for a total of around 6 million people in the mobility dataset. Individual accessibility values are then aggregated at the census tract level based on their home address, to reveal spatial patterns and visualization. 

Census tracts are geographic entities used by the U.S. Census Bureau to collect, analyze, and present its decennial censuses \cite{uscb1994census}.
In general, the population of a census tract is between 2,500 and 8,000, and their boundaries are relatively stable, despite minor updates every ten years due to demographic changes. Our study period (2012 - 2024) includes the 2020 census. To make results and maps consistent, the shape file updated in 2020 \cite{uscb2020tracts} is used to match each individual's home address (represented by longitude and latitude) to the census tract to which it belongs.

For census tract $j$, let $\Is_j$ be the set of individuals whose home is located within the census tract. Then the aggregate value 
$$\overline{\text{TI}}\text{-acs}_j = \frac{1}{\abs{\Is_j}}\sum\nolimits_{i\in \Is_j} \text{TI-acs}_i$$
is the mean of TI-acs$_i$ for all $i \in \Is_j$.
In this study, tract-level TI-acs is aggregated over the resident population rather than conditioning on current EV ownership. Accordingly, it should be interpreted as a place-based measure of charging opportunity under existing mobility patterns, instead of a direct estimate of realized charging experience for current EV drivers.

Most analyses in the \hyperref[sec:results]{Results} section are based on these \emph{tract-level} statistics. Unless otherwise specified, \emph{medians}, \emph{means}, or \emph{quantiles} are according to tract-level statistics, instead of population statistics.

\subsection{Evaluation on unevenness and disparity}\label{sec:uneven_method}
To evaluate the spatial unevenness of TI-acs for residents in different census tracts, plot distributions, and report quantiles, we also compute a \emph{Gini} index. This metric is commonly used to reveal wealth inequality, introduced by Corrado Gini in 1912.

For the discrete case, consider the ``wealth'' distribution $X$ of $N$ entities, whose wealth has been sorted into ascending order as $x_1 \le ... \le x_N$. The Gini index of this distribution is defined as:
\begin{equation}
    \text{Gini}(X) = \frac{\sum_{n=1}^N (2n - N - 1)\, x_n}{N\, \sum_{n=1}^N x_n}
\end{equation}
Gini index is always between 0 and 1, whose two ends respectively represent absolute equality and extreme concentration of resources into one entity.
In our context, the entities are census tracts, and their ``wealth'' is $\overline{\text{TI}}\text{-acs}_j$'s, i.e., the tract-level-average TI-acs of their residents.

In addition, the underlying causes of the revealed disparity is of interest. The correlation between charger accessibility and selected demographic characteristics, particularly race or ethnic group identities, is examined. Accordingly, the coefficient and its confidence interval (CI) are estimated from the following regression:
\begin{equation}
    y_j = \sum_r \beta_r \delta_{jr} + \beta_I \log I_j + \beta_0 + \epsilon_j
\end{equation}
where $y_j$ is the TI-acs metric of census tract $j$, and $\delta_{jr}$ is an indicator variable for whether race/ethnic group $r$ is dominate in the census tract $j$. We include the term $\log I_j$ to control for the effect of income, where $I_j$ is the median household income of the census tract $j$. The identity of the race / ethnic group and the household income statistics are from the American Community Survey (ACS), which is detailed in Section \ref{sec:data}.

\subsection{Data}\label{sec:data}

The evaluation and analysis on TI-acs combines multiple data sources.
Fig.\ref{fig:overview}.\textbf{c} provides an overview of the data processing pipeline, which is detailed in this section.

\paragraph{Mobility dataset by TimeGeo simulation.}
TimeGeo is a mechanistic modeling framework designed to simulate mobility trajectories using passive datasets with mostly sparse traces of individuals \cite{jiang2016timegeo}. 
It employs a time-inhomogeneous Markov chain to model temporal choices and a rank-based exploration and preferential return (r-EPR) model to capture spatial decisions. 
The framework is parameterized by individual-specific parameters, including a weekly home-based tour number, a dwell rate, and a burst rate, supplemented by a global preferential return and exploration rate, and the rank selection probability.
Individual-specific parameters are determined to maximize the likelihood of reproducing observed trajectories from each individual's call detail records (CDR).
In contrast, global parameters are estimated from the aggregated population data, addressing the limitations of sparse individual observations.
More details can be found in the original paper \cite{jiang2016timegeo}.

In this study, the mobility simulation is based on a 6-week long CDR dataset from around 1.39 million users in the Bay area\footnote{While the San Francisco Bay Area is commonly defined to include 9 counties: Alameda, Contra Costa, Marin, Napa, San Mateo, Santa Clara, Solano, Sonoma, and San Francisco, our dataset only contains the lower six, excluding Sonoma, Napa, and Solano.}. Active users with sufficient data for reliable mobility-pattern extraction are first filtered according to predefined criteria; TimeGeo model parameters are then estimated for these users, followed by an expansion procedure to represent the total population within each geographic entity. For each simulated user, their one-week of trajectories are generated by TimeGeo. This expansion preserves important heterogeneity in general mobility behavior through individual-specific TimeGeo parameters and the commuter/non-commuter distinction. However, we should note that it does not explicitly enforce subgroup-level fidelity by race, detailed occupation, or household income.
The TimeGeo framework has been validated with National Household Travel Survey (NHTS) data based on the distributions of stay duration, trip distances, number of visited locations, departure times, mofits, and city-level origin-destination (OD) trips \cite{jiang2016timegeo, xu2018planning}. 
The same simulated Bay area mobility dataset has been used for EV relevant research in \cite{xu2018planning, wu2024planning} among others. 
In this study, we adopt extra postprocessing on the generated trajectories to reflect more accurate stay durations. The main step is to compute the travel time via routing algorithms, which was not explicitly modeled in the original TimeGeo. Detailed procedures are described in \hyperref[sec:appex_B]{Appendix B}.
We refer to the processed dataset as ``TimeGeo data'' hereafter.

\paragraph{Public EVCS dataset.}

Public EVCS information is obtained from the United States Department of Energy's Alternative Fuels Data Center (AFDC) \cite{us_department_of_energy_2024}, accessed in June 2024.
AFDC maintains a repository of EVCS information including locations, charger types and numbers, charging networks, open dates and operation status, accessibility and prices, etc.

The majority of EVCSs recorded in AFDC are public stations, and California state law requires station operators or developers to report data on publicly available chargers to AFDC \cite{cec_ev_chargers_2024}. As of June 2024, there are 4,468 L2 and 1,892 DCFC public charging ports listed on AFDC within our studied area.

Based on the \verb|Open Date| attribute, we create annual ``snapshots'' of charger distributions, i.e. the  subset of EVCSs that were available as of the end of each year (except for 2024, which sets a cutoff in June). We use these snapshots to evaluate charger accessibility in their corresponding years.

To match stays on an individual's travel trajectory with nearby EVCS, we first map the EVCS locations to their nearest nodes on the OSM network. We then create a lookup table for each pair of ``trip stay node'' and ``EVCS node'' that are within 3 km of each other. The distances are computed using shortest-path routing which reflects the actual travel distance. To reduce computational load, the distances are initially filtered using great circle distances. The lookup table allows to efficiently retrieve nearby EVCS stations based on stay nodes when we filter them by multiple conditions, including distance thresholds (0.5, 1, 2, and 3 km) and EVCS open dates.

\paragraph{Census tracts demographics.}

We connect TI-acs with census tract demographic data from the U.S. Census Bureau’s American Community Survey (ACS) 1-year estimates (2012–2022) \cite{us_acs_2024} to analyze their correlations. TI-acs statistics, calculated based on public chargers in operation during a given year, are matched with ACS demographic data from the corresponding year. However, for 2023 and 2024, the latest available ACS data (from 2022) is used due to data availability constraints at the time of the study.
The specific variables used in the analyses herein are summarized in Table \ref{tab:acs} in \hyperref[sec:appex_C]{Appendix C}.

\paragraph{Other data sources \& Software.}
We use Python for data processing, including the use of \verb|geopandas| for geographic data analysis, \verb|OSMnx| \cite{boeing2024modeling}, \verb|NetworkX| \cite{hagberg2008exploring} for routing, and \verb|statsmodels| \cite{seabold2010statsmodels} for statistical analysis.
The regional boundaries in our plots refer to the shapefiles provided by US Census Bureau (TIGER) \cite{uscb2020tracts} and California State Geoportal \cite{ca_geoportal_2024}.

\section{Results}\label{sec:results}

\subsection{TI-acs [hours] across SF Bay area in 2024}

\begin{figure}[h]  % 'h' means here, but LaTeX may move it depending on space
    \centering
    \includegraphics[width=0.9\textwidth]{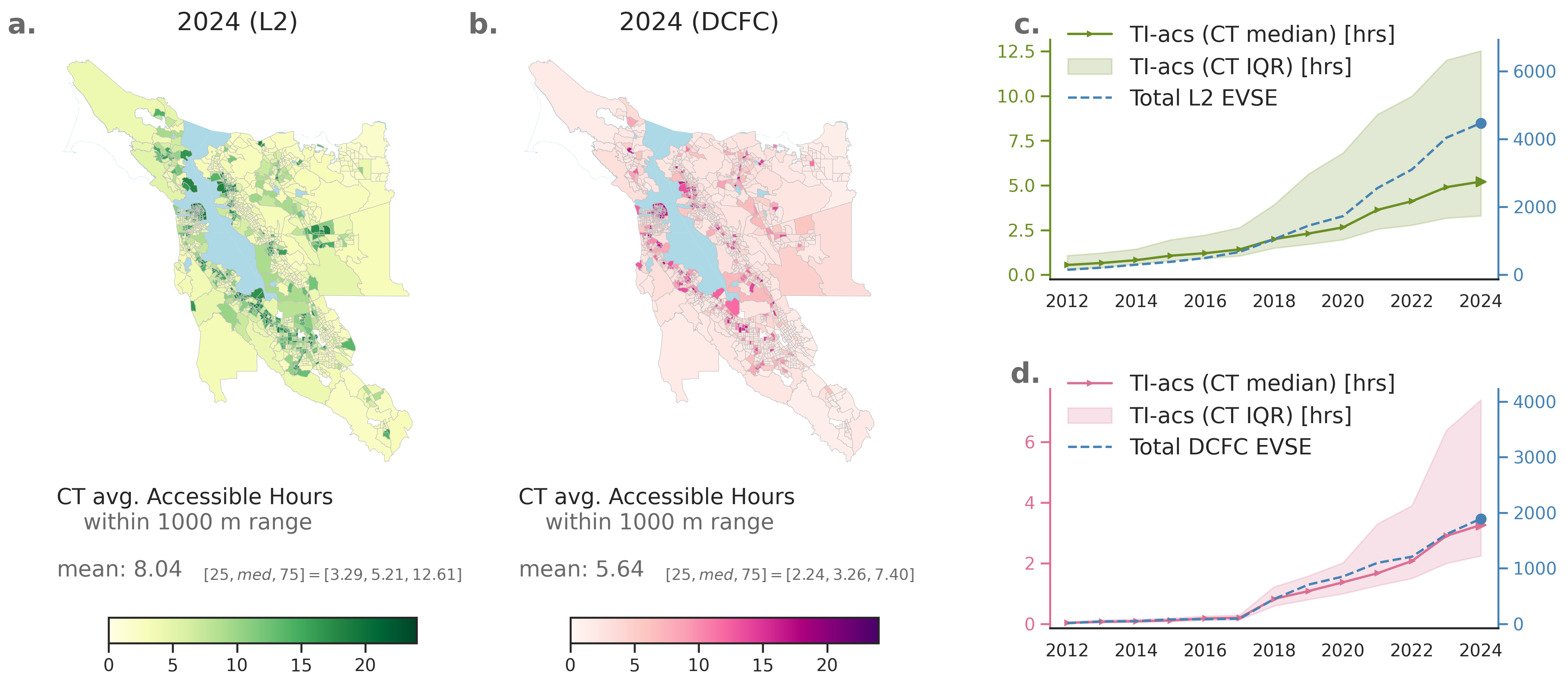}  % Adjust width or use scale instead
    \caption{
    \textbf{Average TI-acs (within 1 km) across Bay area census tracts.}
    {\small\qquad
    \textbf{a},\textbf{b} plot the census-tract-level averaged TI-acs to L2 and DCFC public chargers as of June 2024, respectively.
    Each census tract is colored by the average trajectory-integrated accessible hours (TI-acs) of individuals whose home location is within the area. %(Hatched areas indicate data unavailable.)
    Means, Medians and 1/4, 3/4-quantiles of census-tract-level statistics are annotated.\qquad
    \textbf{c}, \textbf{d} compare the growth of total installed charging ports (blue dashed lines) and the improvement of TI-acs (in terms of accessible hours within 1 km range) from 2012 to 2024.
    Refer to the left $y$-axis for values of TI-acs, and right $y$-axis for values of total installed ports.
    A 1-km distance typically means 10 $\sim$ 15 minutes walk.}
    }
    \label{fig:access_map}  % Label for referencing the figure in the text
\end{figure}

Figure \ref{fig:access_map} shows the average accessible hours (TI-acs [hours]) aggregated by home census tracts in the San Francisco Bay Area.
As of June 2024, the median tract-level-averaged TI-acs across 1428 Bay Area census tracts is 5.21 hours for L2 chargers and 3.26 hours for DCFC. This means that,  during a day, a ``typical'' Bay Area resident has 5.21 hours when at least one public L2 charger is available within 1 km of their location and 3.26 hours when at least one DCFC is available (these two periods may overlap).

Since census tracts vary in population and, more importantly, the distribution of TI-acs across tracts is skewed (more discussion later), the median values deviate notably from the population-weighted averages of TI-acs across all Bay Area residents. The population-weighted averages are 7.50 hours for L2 and 5.17 hours for DCFC, both significantly higher than the medians, indicating a spatially uneven distribution of public charging accessibility assessed via TI-acs.

As of June 2024, the top 25\% of census tracts, which are home to approximately 1.2 million Bay Area residents, have a tract-level-averaged TI-acs above 12.5 hours for L2 and 7.39 hours for DCFC. In contrast, the bottom 25\% of census tracts, covering 1.5 million residents, have TI-acs below 3.29 hours for L2 and 2.24 hours for DCFC. 
The nearly fourfold difference between the 1st and 3rd quartiles highlights the considerable spatial variation in people’s daily experience with public charging infrastructure.

\subsection{Location-based TI-acs breakdown}

The metric accessible hours (TI-acs [hours]) is \emph{additive} when evaluated across different segments. This property allows for a straightforward breakdown, providing deeper insights into accessibility across various locations (activity types) and different times of the day.

\begin{figure}[h!]  % 'h' means here, but LaTeX may move it depending on space
    \centering
    \includegraphics[width=0.98\textwidth]{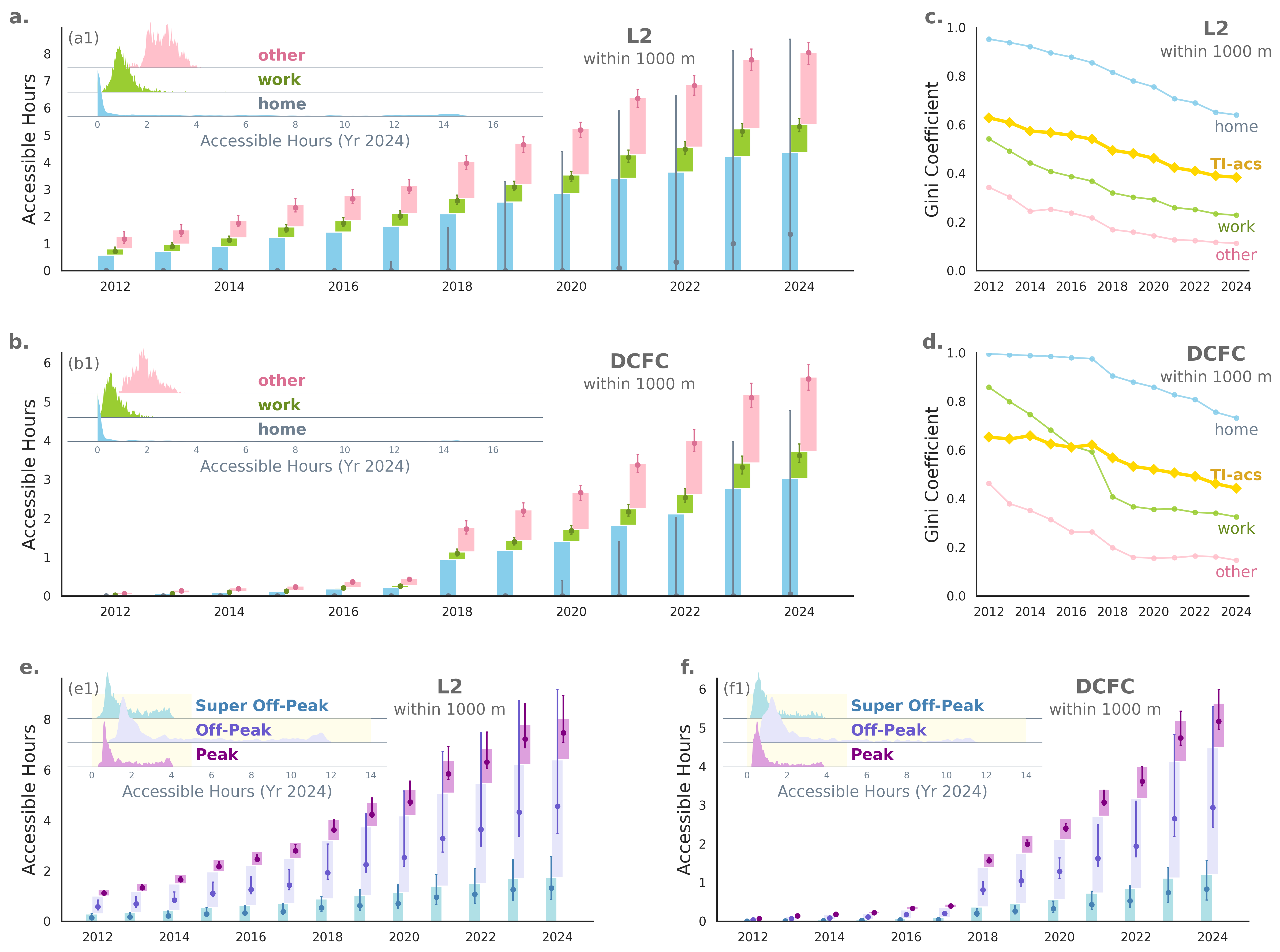}  % Adjust width or use scale instead
    \caption{\textbf{TI-acs breakdown.}
    {\small\qquad
    \textbf{a},\textbf{b} plot the breakdown of TI-acs (accessible hours) into different stay location segments: home, work, other by different years from 2012 to 2024, for L2 and DCFC respectively.
    Height of bars represent the means of census-tract-averaged metric, and the darker error bars annotate quartiles (25\%, 50\%, 75\%) of the metric (with $0$ from the bottom of corresponding bars). 
    Subplots (a1), (b1) are density of accessible hours in different segments across all census tracts in year 2024.\qquad
    \textbf{c}, \textbf{d} plot the annual change of Gini index of TI-acs, as well as its breakdown into different location segments from 2012 to 2024, for L2 and DCFC respectively.
    The higher Gini index is, the more uneven charger accessibility is for people reside in different census tracts.\qquad
    \textbf{e}, \textbf{f} plot the breakdown of TI-acs into different time segments, defined by power grid TOU periods, by different years from 2012 to 2024, for L2 and DCFC respectively, with
    subplots (e1), (f1) visualizing the density in year 2024.
    The meanings of different plot elements are the same as they are in \textbf{a}, \textbf{b}.
    The total duration of different TOU periods in a day are: super off-peak: 5 hr, off-peak: 14 hr, peak: 5 hr, as indicated by the light yellow shades in (e1), (f1).
    }}
    \label{fig:breakdown}  % Label for referencing the figure in the text
\end{figure}

As of June 2024, in terms of mean values, individuals, on average, have a greater share of accessible hours while at home compared to when they are at workplaces or other venues. This is primarily due to the extended duration of home stays, including approximately one-third of the day spent sleeping. However, public charger accessibility near homes exhibits a pronounced long-tail distribution, making the mean value less representative of the broader population. As shown in the distribution plot (Fig. \ref{fig:breakdown}.~\textbf{a1}, \textbf{b1}), a substantial proportion of census tracts have residents with little to no access to public EV chargers near their homes. Given that individuals typically spend around 12 hours per day at home, a census tract with an average home-based accessible hour as low as 1 hour (which applies to 49\% (L2) / 59\% (DCFC) of census tracts in 2024) implies that fewer than 10\% of its residents have at least one public charger within 1 km of their homes. This limited accessibility stems from multiple factors, including zoning policies, as public chargers are predominantly installed in commercial and industrial areas, while residential zones rely partially on privately owned home chargers. Nevertheless, much of the previous literature has focused solely on public charger accessibility near homes, making it an incomplete metric for evaluating both effectiveness and equity.

In contrast, the distribution of accessible hours at workplaces and other venues is closer to a normal distribution with lower variance. This can be attributed to greater population mixing in these locations, where the same charging infrastructure is shared among a diverse set of individuals from different neighborhoods. In terms of median values, as of June 2024, accessible hours at ``other’’ locations (2.6 hours) exceed those at home (1.4 hours) for L2 chargers and are significantly higher for DCFC chargers (other: 1.9 hours, home: 0.0 hours). Since DCFC chargers are almost exclusively located in commercial and industrial areas, this mixing effect becomes even more pronounced. When interpreting workplace accessibility, it is important to note that approximately 40\% of the population, classified as ``non-commuters’’, do not have a regular workplace.

\subsection{Time-of-use (TOU) -based TI-acs breakdown}

A key consideration in sustainable EV charging is the flexibility to shift charging demand from peak hours to off-peak periods. Such a shift can reduce peak demand, enhance grid stability and resilience, while also reducing carbon emissions because off-peak hours generally coincide with a cleaner energy mix. However, for this demand shift to be feasible, a fundamental prerequisite is that EVs have access to chargers during off-peak times.

To quantify this \emph{potential}, we segment the day into super off-peak (9 AM – 2 PM), off-peak (9 PM – 9 AM, 2 PM – 4 PM), and peak (4 PM – 9 PM) periods, following the PG\&E business EV rate plan. Then we assess EV drivers’ accessible hours (TI-acs [hours]) within these specific time windows. 
As illustrated in Fig. \ref{fig:breakdown}.~\textbf{e},\textbf{f}, the median accessible hours for Bay Area residents as of June 2024 are 1.1 hours during a total of 5 peak hours, 2.8 hours during 14 off-peak hours, and 1.3 hours during 5 super off-peak hours per day for public L2 chargers. For DCFC chargers, the corresponding median accessible hours are 0.7, 1.7, and 0.8 hours, respectively.

These findings indicate that individuals generally have greater access to public charging infrastructure (particularly L2 chargers) during super off-peak hours compared to peak hours. This is largely due to people’s mobility patterns where they are more likely to be at workplaces or public venues in the morning and early afternoon super off-peak period, where access to public chargers is more prevalent. This insight highlights a promising opportunity to encourage more EV charging during periods when the grid is powered by cleaner energy. However, to fully realize this potential, policy and market mechanisms must be carefully designed to incentivize and facilitate off-peak charging adoption.

\subsection{Temporal evolution over the past decade}

Over the past decade, rapid advancements in technology and significant market growth have transformed the EV and charging infrastructure landscape. Key indicators, such as the number of public chargers, have increased exponentially. In 2018, the Bay Area had only about 1,000 public L2 chargers and fewer than 500 DCFC units. By June 2024, these figures had risen to approximately 4,500 public L2 chargers and 2,000 DCFC units. Despite disruptions caused by the pandemic, the total number of installed public L2 and DCFC chargers grew by 350\% and 300\%, respectively, in less than six years.

During the same period, median census-tract-averaged accessible hours (TI-acs [hours]) for public L2 and DCFC chargers increased from 2.0 hours and 0.82 hours in 2018 to 5.2 and 3.2 hours as of June 2024, representing 180\% and 290\% growth, respectively. 
While these improvements, particularly for L2 chargers, are substantial in absolute terms, they lag behind the overall increase in charger installations. This discrepancy arises because TI-acs [hours] does not account for the number of chargers at a given location but rather reflects the spatial distribution of charging infrastructure. The slower growth in TI-acs [hours] relative to total charger installations suggests that new L2 chargers have not been deployed in locations that would maximize improvements in charging accessibility for communities with historically low accessibility.

Trends in TI-acs [hours] are shown in Fig.~\ref{fig:breakdown}, alongside changes in the Gini index measuring disparities in public charger accessibility across locations (Fig.~\ref{fig:breakdown}\textbf{c},\textbf{d}). A Gini index of 0 indicates perfect equality, whereas a value of 1 represents extreme concentration of resources. As infrastructure has expanded, accessibility inequality has steadily declined across all segments, as reflected by decreasing Gini index. By 2024, the Gini index for TI-acs [hours] was 0.38 for L2 chargers and 0.44 for DCFC chargers, marking reductions of 24\% and 23\%, respectively, since 2018. 
The larger disparity for DCFC suggests that fast-charging accessibility is more spatially concentrated than L2 accessibility. One likely reason is infrastructure placement: L2 chargers are deployed across a broader mix of settings, while DCFC stations are more selectively sited in commercial and other high-activity corridors rather than near residential areas. 
Another notable pattern is the high concentration of accessibility near homes, whereas accessibility at workplaces and other public locations is much more evenly distributed.

\subsection{Racial disparity on TI-acs}\label{sec:res_race}

The substantial spatial variance in TI-acs raises a critical question regarding the equity and inclusiveness of public charger accessibility. In particular, we investigate whether there is evidence of racial disparities in TI-acs.

\begin{figure}[h!]  % 'h' means here, but LaTeX may move it depending on space
    \centering
    \includegraphics[width=\textwidth]{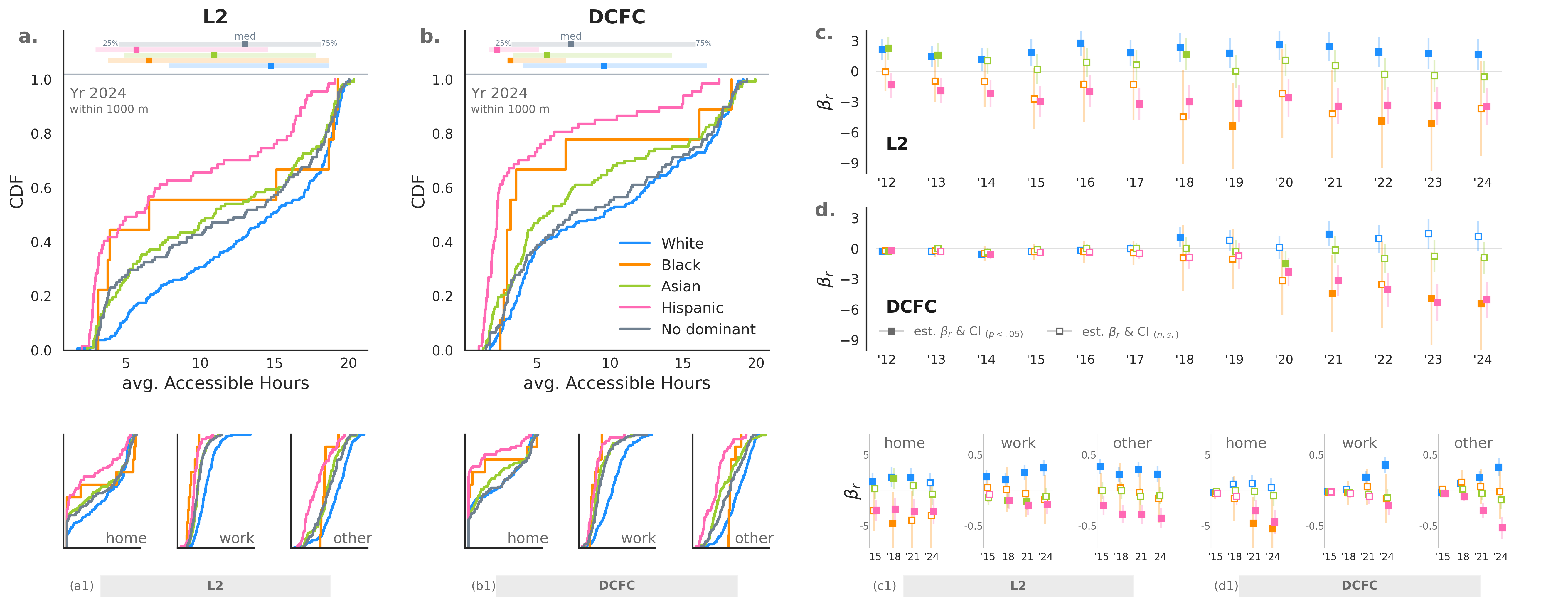}  % Adjust width or use scale instead
    \caption{\textbf{Racial disparities on TI-acs}
    {\small\qquad
    \textbf{a}, \textbf{b} plot the cumulative density function (CDF) of census-tract-level-averaged TI-acs (accessible hours) in year 2024 
    grouped by the dominate race (ethnic identity) of the census tracts,  for public L2 and DCFC, respectively. 
    Horizontal bars and markers at the top mark the quartiles (25\%, 50\%, 75\%) of TI-acs for different groups, sharing the same $x$-axis with the main plots.
    Subplots (a1), (b1) plot the CDF of accessible hours in different stay location segments in year 2024.\qquad
    \textbf{c}, \textbf{d} visualize the coefficients $\beta_r$ from the regression: $y_i = \sum_r \beta_r \delta_{ir} + \beta_I \log I_i + \beta_0 + \epsilon_i$.
    For census tract $i$, $y_i$ is its accessibility metric, $\delta_{ir} = 1$ if it is in group $r$ (white, black, Asian, Hispanic), otherwise $0$, and $I_i$ is its median household income.
    $\beta_r$ (unit: hours) means the difference in daily accessible hours between people residing in the census tracts with a given dominate race and people in census tracts with no dominate race.
    We ran the regression for each year in 2012 to 2024 separately. 
    Markers are the ordinary least square (OLS) estimate of $\beta_r$'s, and the error bars visualize the confident interval (CI) of the corresponding estimate. 
    A solid marker indicates that the coefficient is significantly ($p<.05$, double side) greater/smaller than 0. 
    \qquad For all results plotted, we only include census tracts which more than half of its households live in 2-unit (or more) structures.
    }
    }
    \label{fig:race}  % Label for referencing the figure in the text
\end{figure}

To examine this, we plot the \emph{cumulative density function} (CDF) of tract-level TI-acs, grouping census tracts by their dominant racial group.\footnote{``Hispanic'' is a cultural and ethnic identity rather than a biological classification. Hispanic populations are racially diverse. In most contexts, when we use the term ``race'', we refer to ``race or ethnic group''.} A racial group is identified as \emph{dominant} in a census tract if it constitutes both the largest racial group and at least 40\% of the total population. The CDF plot serves as an intuitive illustration of racial disparities: curves positioned toward the upper left indicate a higher concentration of census tracts with lower TI-acs, whereas curves toward the lower right indicate more tracts with higher TI-acs.
Since our EVCS dataset does not include private chargers, we mitigate potential bias by conducting analysis on a subset of census tracts where more than 50\% of residents live in multi-unit dwellings (MUDs), as MUD residents are generally less likely to have access to private home chargers compared to those in detached houses. Besides, these disparities should be interpreted as differences in the public-charging opportunity environment facing residents of different communities, rather than as direct estimates of charging outcomes among current EV owners only.

As shown in Fig. \ref{fig:race} \textbf{a},\textbf{b}, the CDF curve for Hispanic-majority communities appears in the upper left region, indicating that many of these communities experience lower TI-acs. In contrast, White-majority communities are positioned toward the lower right, suggesting higher TI-acs. While previous studies on charging accessibility have reported racial disparities, most of them rely on home-based metrics. Our analysis reveals that these disparities persist across all activity segments, including home, workplaces, and other locations (Fig. \ref{fig:race} \textbf{a1},\textbf{b1}).

Regression analysis of TI-acs against racial dominance further supports the observed disparities, as illustrated in Fig. \ref{fig:race}. Using the method detailed in Sec. \ref{sec:uneven_method}, the results show that in 2024, compared to census tracts with no dominant racial group, White-dominated census tracts have 1.68 more accessible hours per day for public L2 chargers {\scriptsize (95\% CI: [0.18, 3.18])} and 1.19 additional hours for public DCFC chargers {\scriptsize (95\% CI: [-0.27, 2.65])}. In contrast, Hispanic-dominated census tracts experience 3.44 fewer accessible hours for L2 chargers {\scriptsize (95\% CI: [-5.29, -1.59])} and 5.07 fewer hours for DCFC chargers {\scriptsize (95\% CI: [-6.87, -3.28])}.

As shown in Fig. \ref{fig:race} \textbf{c1}, \textbf{d1}, residents in Hispanic-dominated census tracts consistently have lower TI-acs for both L2 and DCFC chargers across all activity segments - home, workplace, and other locations - compared to residents in White-dominated census tracts. Although the differences are most pronounced near their homes, the disparities at work and other locations of daily activity are also statistically significant. This pattern suggests segregated mobility patterns reflected by racial disparities in work locations (due to occupational distribution) and other activity spaces (shaped by economic and cultural factors), which are accompanied by an uneven distribution of public charging stations.

\section{Discussions}

In this section, we present additional analyses and visualizations related to TI-acs assessment and racial disparities in public EV charging accessibility. 
We then explore the broader implications of our findings, situating them within the context of existing research. 
Finally, we discuss the limitations of our study and outline potential directions for future work.

\subsection{Accessibility by Different Measures, Distance Thresholds, and Years}

\begin{figure}[h!]  % 'h' means here, but LaTeX may move it depending on space
    \centering
    \includegraphics[width=\textwidth]{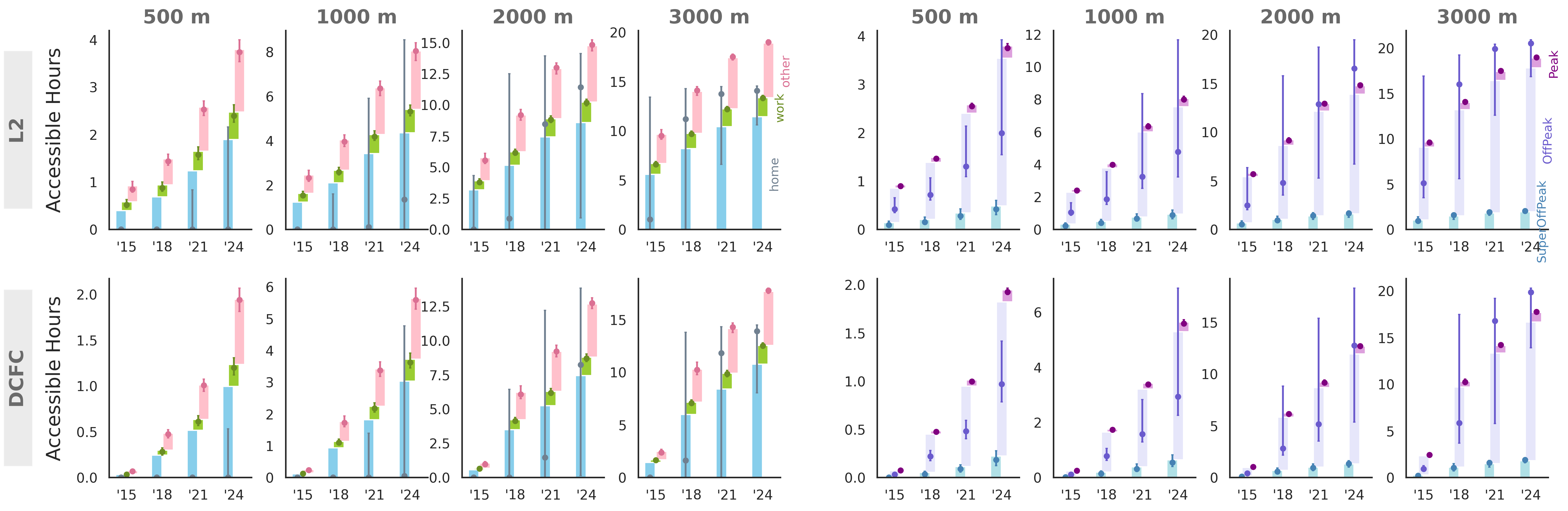}  % Adjust width or use scale instead
    \caption{\textbf{TI-acs [hours] breakdown by different distance thresholds}
    {\small\qquad
    Subplot titles specify the accessible radius. $x$-axis marks EVCS snapshots in different years. \emph{top row}: L2, \emph{bottom row}: DCFC. \emph{left half}: breakdown by stay location types. \emph{right half}: breakdown by time in the day. 
    \qquad Please refer to Fig.\ref{fig:breakdown} for detailed legends.
    }
    }
    \label{fig:sensitivity}  % Label for referencing the figure in the text
\end{figure}

We vary the ``accessible radius'' (i.e., the maximum distance within which EV chargers are considered accessible) among 500, 1000, 2000, and 3000 meters and evaluate TI-acs [hours] (and its breakdown) accordingly, as shown in Fig. \ref{fig:sensitivity}. As expected, accessible hours increase with radius: expanding the threshold from 500 m to 1 km nearly doubles accessible hours, whereas increasing it from 2 km to 3 km yields less than a 20\% gain, indicating diminishing returns. Importantly, the temporal trends and qualitative spatial patterns remain consistent across thresholds, suggesting that our main findings are robust to this modeling choice.
We use 1 km as the primary threshold because it corresponds roughly to a 12 - 15 minute walk and provides a transparent middle ground between overly restrictive and overly permissive definitions of destination-based charging access. This choice is also informed by data limitations \cite{jiang2016timegeo, xu2018planning}: the spatial resolution of the TimeGeo locations is approximately 200 - 300 m, and additional error is introduced when locations are matched to the nearest road-network nodes, making substantially smaller thresholds more sensitive to measurement noise.

\begin{figure}[h!]  % 'h' means here, but LaTeX may move it depending on space
    \centering
    \includegraphics[width=\textwidth]{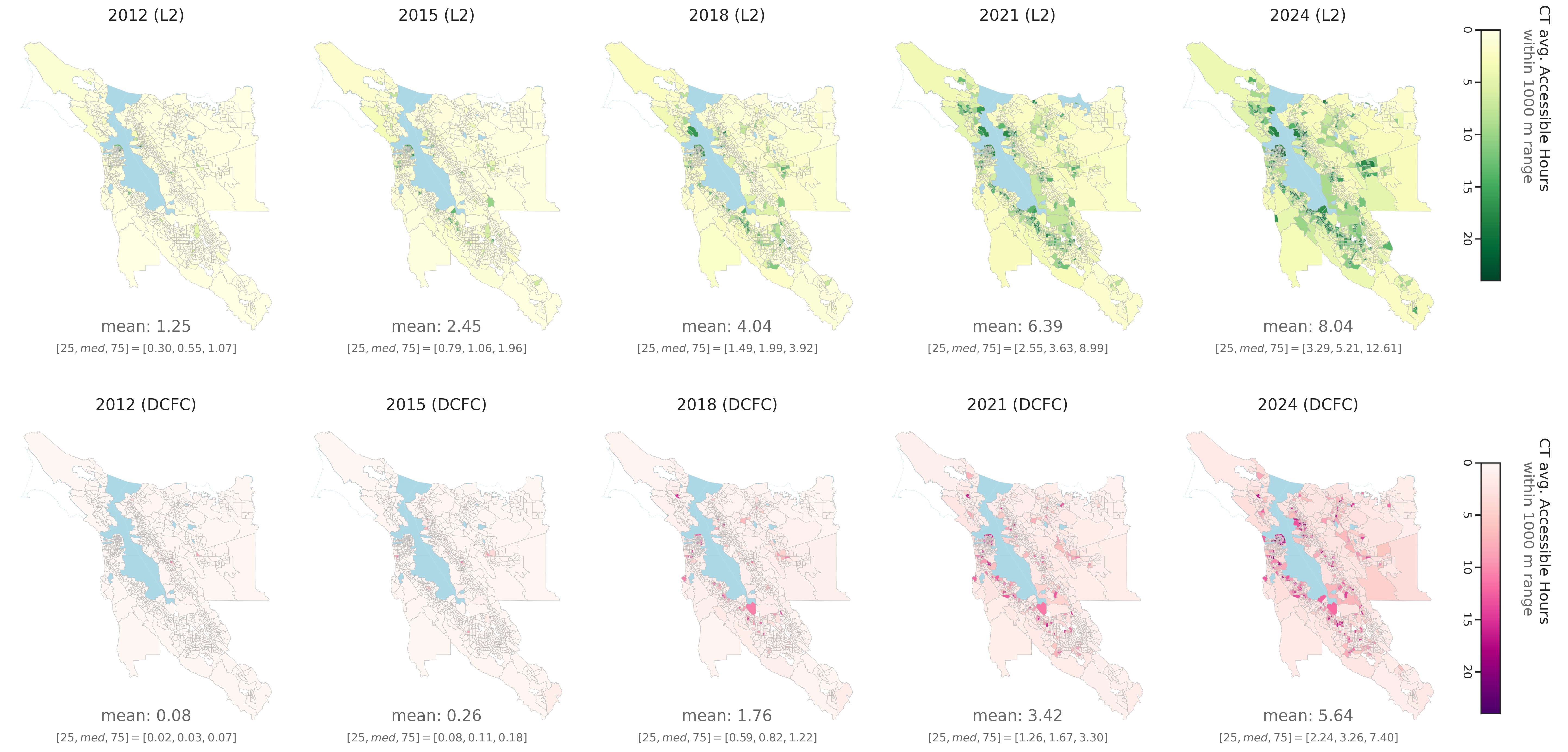}  % Adjust width or use scale instead
    \caption{\textbf{TI-acs [hours] map across Bay Area census tracts in multiple years}
    % {\small\qquad
    % XXX
    % }
    }
    \label{fig:TI-acs-hour-year}  % Label for referencing the figure in the text
\end{figure}

\begin{figure}[h!]  % 'h' means here, but LaTeX may move it depending on space
    \centering
    \includegraphics[width=\textwidth]{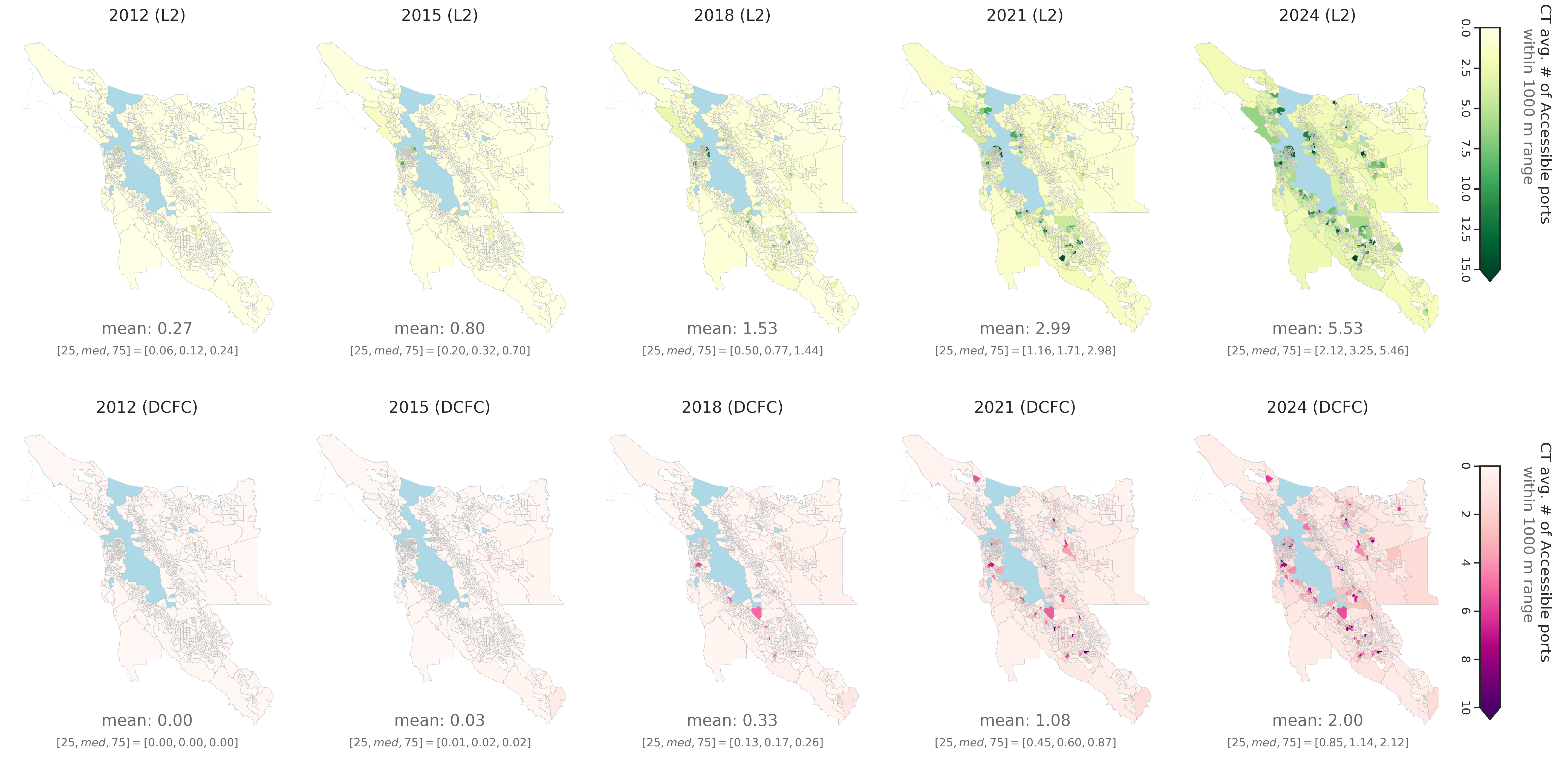}  % Adjust width or use scale instead
    \caption{\textbf{TI-acs [ports] map across Bay Area census tracts in multiple years}
    % {\small\qquad
    % XXX
    % }
    }
    \label{fig:TI-acs-port-year}  % Label for referencing the figure in the text
\end{figure}

In Fig. \ref{fig:TI-acs-hour-year} (\hyperref[sec:appex_A]{Appendix A}), we present maps of census tract–level TI-acs [hours] across multiple years, illustrating the temporal and spatial evolution of public charging accessibility. We also visualize the corresponding trends for the TI-ACS [ports] metric in Fig. \ref{fig:TI-acs-port-year}, which represents the average number of public charging ports (separately for L2 and DCFC) encountered during stays along individuals’ trajectories. For L2 chargers, the mean value at the census tract level increased from 0.27 in 2012 to 1.53 in 2018 and 5.53 in 2024. For DCFC, the corresponding mean increased from 0.00 in 2012 to 0.33 in 2018 and 2.00 in 2024. These trends closely align with those observed for TI-acs [hours], which are analyzed in detail in Section XX.

In this study, we adopt TI-acs [hours] (\textit{a.k.a.} accessible hours) as the primary metric for two main reasons. First, by combining typical charging power with vehicle energy efficiency, accessible hours can be interpreted as a rough estimate of additional driving range enabled by opportunistic charging. Although this interpretation may be optimistic for neglecting the competition at public chargers, it provides a convenient indicator of electrification potential, especially in regions with persistently low accessibility. Second, in areas with already high charger density, additional deployment yields diminishing returns when measured in terms of accessible hours, aligning more closely with how EV drivers perceive real-world charging convenience.
That said, no single metric fully captures the complexity of public charging accessibility. Instead, complementary measures serve to enrich our understanding and provide a more comprehensive view of the evolving infrastructure landscape.

\subsection{Robustness test \& discussion on racial disparities}

\begin{figure}[h!]  % 'h' means here, but LaTeX may move it depending on space
    \centering
    \includegraphics[width=\textwidth]{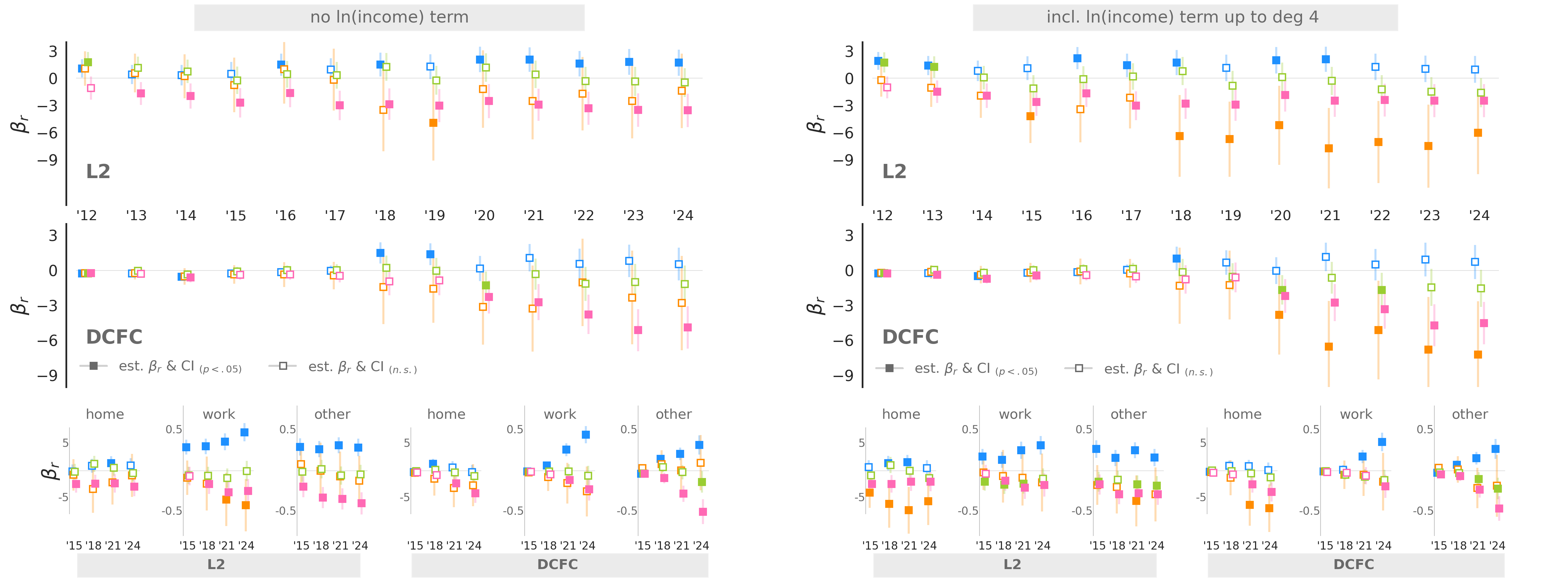}  % Adjust width or use scale instead
    \caption{\textbf{Statistical test Racial disparities on TI-acs}
    {\small\qquad
    \emph{left}: no income term, i.e., $\beta_r$'s from regression
    $y_i = \sum_r \beta_r \delta_{ir} + \beta_0 + \epsilon_i$.
    \emph{right}: income (logarithm) term up to degree 4, i.e., $\beta_r$'s from regression
    $y_i = \sum_r \beta_r \delta_{ir} +
    \sum_{k=1}^4 \beta_I^k (\log I_j)^k + \beta_0 + \epsilon_i$.\qquad
    Please refer to Fig. \ref{fig:race} for detailed legend.
    }
    }
    \label{fig:race-reg-appx}  % Label for referencing the figure in the text
\end{figure}

Our analysis of the relationship between racial composition and TI-acs is observational and correlational. While it reveals that residents in Hispanic-dominated communities have significantly lower TI-acs than those in White-dominated communities, the available data do not further allow us to pinpoint the specific mechanisms driving this disparity, and it is likely the result of multiple interrelated factors.

Racial composition in census tracts is historically correlated with economic status, which in turn influences both the deployment of public charging infrastructure in the living neighborhoods and residents’ exposure to locations outside. In Sec. \ref{sec:res_race}, we attempt to disentangle the effects of economic status from other less explicit and well-understood factors by including a control term, $\log I_j$, in our regression model, where 
$I_j$ represents the median household income in the census tract $j$. Fig. \ref{fig:race-reg-appx} presents additional regression results both without this control term (left) and with a higher-order (4th-degree) income term (right). These results remain consistent with our main findings from Sec. \ref{sec:res_race}.

Introducing higher-order income terms tends to reduce the estimated coefficients for dominant race indicators, as a more flexible income term can explain a larger share of the variance in the data. However, even after allowing income to absorb as much variance as possible, Hispanic-dominated census tracts still exhibit significantly lower TI-acs. This suggests that the disparity is also driven by other structural or systemic non-economic factors, such as historical patterns of infrastructure investment, differences in mobility behaviors, preferences in vehicle fuel types, and the priority given to these communities in public decision-making processes.

Since the TimeGeo dataset does not include explicit demographic labels, these results should be interpreted as tract-level disparities in charging opportunity experienced by communities under the simulated mobility field, rather than as proof that every subgroup-specific mobility mechanism is fully recovered at the individual level.

\subsection{Impact of within-tract aggregation metric}
\label{sec:agg_metric}

Our main analysis summarizes individual TI-acs within each census tract using the tract-level mean, which can be interpreted as the expected accessibility experienced by a randomly selected resident of the tract. However, because individual TI-acs within a tract may be skewed due to heterogeneous daily activity spaces, it is also informative to examine alternative within-tract summaries that emphasize the worse-served, typical, or better-served portions of the tract population.

Table~\ref{tab:agg_metric_compare} reports tract-level summary statistics in 2024 when the within-tract aggregation is taken as the mean, 25th percentile, or 75th percentile, for both L2 and DCFC TI-acs under the 1-km threshold. 

\begin{table}[h!]
\centering
\caption{Tract-level TI-acs [hours] statistics in 2024 under alternative within-tract aggregation metrics. 
\scriptsize \textbf{Rows} indicate \textbf{within-tract} aggregation metric, i.e., how individual TI-acs values are aggregated within each census tract; \textbf{Columns} report \textbf{cross-tract} statistics, i.e., the corresponding distribution of tract-level values across all census tracts.}
\label{tab:agg_metric_compare}
\normalsize
{%
\begin{tabular}{@{}l | cccc | cccc@{}}
\toprule
& \multicolumn{4}{c|}{\textbf{L2 (1 km)}} & \multicolumn{4}{c}{\textbf{DCFC (1 km)}} \\ \cmidrule(lr){2-5} \cmidrule(l){6-9}
\textbf{metric} & \textbf{mean} & \textbf{25\%} & \textbf{med} & \textbf{75\%} & \textbf{mean} & \textbf{25\%} & \textbf{med} & \textbf{75\%} \\ 
\midrule
\textbf{mean}  & 8.04  & 3.29 & 5.21 & 12.61 & 5.64 & 2.24 & 3.26 & 7.40  \\
\textbf{25\%}  & 4.09  & 0.45 & 1.16 & 4.66  & 2.17 & 0.02 & 0.26 & 1.02  \\
\textbf{75\%}  & 11.50 & 5.14 & 7.43 & 19.85 & 8.27 & 3.39 & 4.93 & 14.10 \\ \bottomrule
\end{tabular}%
}
\end{table}

The main qualitative findings are preserved across aggregation choices. In all cases, TI-acs increases substantially over time for both L2 and DCFC; L2 remains more accessible than DCFC in absolute terms; and DCFC remains more spatially unequal than L2. The broad decomposition by stay type is also unchanged: accessibility associated with workplaces and other non-home locations is much more evenly distributed than accessibility near home, and overall inequality declines as charging infrastructure expands. Likewise, the main spatial and demographic disparity patterns remain stable. Using the 75th percentile instead of the mean shifts tract-level TI-acs upward mechanically, but does not alter the broad ranking of better- and worse-served areas or the direction of disparities across tract groups.

At the same time, alternative aggregation metrics reveal several useful nuances. First, when the 75th percentile is used, DCFC TI-acs grows more rapidly than installed DCFC EVSE, whereas under mean-based aggregation the opposite pattern is observed. This suggests that early DCFC expansion disproportionately improved accessibility for the better-served portion of residents within each tract before diffusing more broadly.
Second, inequality in workplace accessibility appears substantially stronger under the 75th-percentile aggregation than under the mean, especially for L2. This suggests that the better-served residents within each tract are concentrated in a relatively small number of tracts with strong workplace access, a pattern that is partially masked by tract averages.

\subsection{Relevant research in adjacent fields}
It is worth noting that metrics similar to TI-acs have been proposed in other application domains beyond EV charging. For example, Testi \textit{et al.} \cite{testi2024big} combined mobility data with hyperlocal PM2.5 measurements in the Bronx, New York, to assess disparities in air pollution exposure across different population groups. Nevertheless, residence-based accessibility metrics remain the dominant approach in most fields when evaluating access to various urban services \cite{wang2021access}, which appeals to new research directions through a mobility lens \cite{xu2025using, yabe2024enhancing}.

\subsection{Limitations \& future work}

\paragraph{Distance threshold in TI-acs}
Our analysis uses a fixed 1-km radius as the primary definition of accessibility, which provides a transparent and empirically supported baseline. However, the appropriate access threshold is likely context-dependent: a 1-km walk may be reasonable for a multi-hour workplace stay, but much less so for a short errand or time-sensitive stop. More generally, charging accessibility may be better represented by a distance-decay or tolerance spectrum rather than a binary cutoff. Future work could therefore extend TI-acs using context-specific thresholds or weighted formulations calibrated against observed charging choices and user experience. We view TI-acs not as a single fixed metric, but as a flexible framework that can accommodate such behavioral refinements. In addition, note that this walking-based threshold is most natural in dense urban settings. In less dense, more car-dependent regions, accessibility may be better characterized by detour-based measures, such as additional driving time or route deviation required to reach a charger.
The present metric does not account for the additional detour or routing burden required to reach a charger within a trip chain, which may be especially relevant for short stays or for chargers located in large commercial parcels.

\paragraph{EV ownership}
It is well known that the demographics of current EV drivers differ significantly from those of the general population. In this study, we compute TI-acs over all residents in a census tract rather than conditioning on current EV ownership. As a result, our tract-level averages should not be interpreted strictly as the realized charging experience of current EV drivers today. Instead, they primarily characterize the accessibility environment that residents of a tract would face if they were to adopt an EV, making TI-acs particularly useful as a forward-looking planning and equity metric. This perspective is important because communities with historically lower infrastructure investment may also have lower EV adoption today; excluding them would risk understating inequities in the broader electrification transition. As a practical constraint, our mobility dataset does not include EV-driver identifiers, and existing Bayesian approaches to infer EV ownership have not yet been validated at the individual level. Future work with linked vehicle-ownership and mobility data could better distinguish present-day EV-driver experience from future accessibility potential.

\paragraph{Private chargers}
Our study focuses on accessibility to public EV chargers, but \emph{shared private chargers} (available only to specific groups, such as those in workplaces or multifamily housing complexes) and \emph{private home chargers} also play crucial roles in the U.S. charging infrastructure. It is estimated that the number of shared private L2 chargers is approximately 1.5 times that of public L2 chargers \cite{cec_ev_chargers_2024}, and private L2 home chargers account for around 80\% of total installed chargers in the U.S. as of 2020 \cite{bauer2021charging}, with approximately 70\% of charging sessions occurring at home today. (Most DCFC chargers are public.) Unfortunately, these estimates, derived from county-level surveys conducted by charging service providers, lack the granularity needed for direct integration into our TI-acs calculations.
Additionally, while the AFDC dataset is widely used in EVCS-related research, it may not capture 100\% of public chargers, even in California. A recent update from the California Energy Commission (August 2024) incorporated additional public chargers identified through PlugShare data \cite{cec_ev_chargers_2024}, though detailed information on these chargers remains unavailable.

The omission of private chargers is a significant limitation of our study (as well as many other related studies). 
In particular, the limitation may be more consequential in suburban and rural regions, where detached housing and home charger installation are more prevalent and public charging may play a smaller role in routine charging access.
We attempt to mitigate potential biases by conditioning some of our analyses on land use — specifically, focusing on MUD-dominant communities, as discussed in Sec. \ref{sec:res_race}. A relevant research direction is: For potential EV adopters, how likely are they to install a home charger when they adopt an EV? Researchers interested in better estimating future home charger installations may refer to methods outlined in \cite{brockway2021inequitable}, which integrate land use data with distribution grid capacity constraints.

\paragraph{Realized charging feasibility}
Even when a public charger is geographically accessible, actual charging feasibility depends on additional factors beyond proximity, including detour cost within a trip chain, compatibility with the available dwell time at the stop, and competition for the charger.TI-acs metric already captures dwell-time opportunity at a first-order level, but it does not model route deviation or queueing explicitly. 

Even for public chargers, their use is exclusive in the sense that each charger can serve only one EV at a time. As a result, our TI-acs estimates likely overstate actual accessibility since they do not account for congestion at charging stations. With currently available data and models, we can estimate the number of individuals near each charger at any given time, which could provide a rough indication of congestion. However, such estimates remain far from the actual situation, as not all individuals in proximity require charging, nor do they utilize every available charger nearby.
Addressing this limitation requires integrating behavioral models of EV drivers’ charging decisions into the analysis, adding a significant layer of complexity that we defer to future research. Existing EV driver behavioral models are also insufficient — many rely on simple rule-based approaches fitted to historical (and often outdated) data, making them unreliable for extrapolation. Moreover, they fail to capture strategic planning capability of EV drivers over long time horizons and lack the flexibility to incorporate many charging behavior patterns. Emerging agent-based simulation techniques, such as large language model (LLM)-powered agents, present promising opportunities to tackle these challenges. However, their reliability and scalability remain open questions, requiring further validation before they can be effectively applied in this context.

\paragraph{TI-acs vs EV adoption}
Finally, TI-acs offers a novel perspective on charging resource availability, which could provide fresh insights into how charging accessibility influences EV adoption. However, establishing causality remains challenging. The well-known ``chicken-and-egg’’ dilemma—where charging infrastructure and EV adoption influence each other—complicates the analysis. While we hypothesize that greater accessibility reduces charging anxiety and encourages EV adoption, it is difficult to disentangle this effect from the tendency of charging service providers to install more chargers in areas with higher EV adoption. Additional challenges include the availability of high-resolution EV adoption data. For example, the CVRP dataset \cite{cvrp_2024}, used in several EV relevant studies \cite{xu2018planning, wu2024planning, lee2019buying}, does not include Tesla buyer or sales from used vehicle markets. Moreover, it reports sales at Zip code level, which does not match with the geographic granularity of our analysis. Therefore, we leave the investigation to future study.

\paragraph{Transferability across regions} 
The present case study focuses on the San Francisco Bay Area, where a walking-based accessibility threshold is a reasonable first-order approximation for many daily activities. In less dense, more car-dependent regions, this assumption may be less realistic, and public-charger accessibility may be better represented by detour-based measures such as added driving time or route deviation. In addition, private home charging is likely to play a larger role in such settings. We therefore view TI-acs as a flexible trajectory-based framework whose implementation should be adapted to regional context, rather than as a single fixed metric calibrated universally across urban, suburban, and rural environments.

\subsection{Policy implications}
The TI-acs framework has direct relevance for public agencies and utilities seeking to accelerate equitable and grid-compatible deployment of public EV charging infrastructure. 
TI-acs can be converted into \emph{deployment priority maps} that guide infrastructure investment toward communities with the greatest accessibility deficits. 
Importantly, as TI-acs is mobility-informed, these priority maps do not reduce to residential proximity alone; rather, they can highlight under-served commercial corridors, employment centers, and high-dwell activity clusters where public charging opportunity is presently limited despite substantial daily presence. 
Such maps can be operationalized in planning workflows by (i) selecting a policy threshold (e.g., bottom quartile of tract-level TI-acs), (ii) identifying candidate parcels and public right-of-way opportunities within those tracts, and (iii) screening candidates by feasibility constraints (grid capacity, permitting, and co-location with amenities) before proceeding to detailed site design.

Second, the persistence of spatial inequality in TI-acs motivates stronger \emph{equity guardrails} in infrastructure programs. In the U.S., federal and state programs such as the National Electric Vehicle Infrastructure (NEVI) Formula Program and related public funding streams explicitly recognize the need to serve underserved or disadvantaged communities. TI-acs provides an implementable metric for translating such equity intent into quantitative targeting criteria, for example by directing a portion of program funds toward tracts where TI-acs falls below a defined threshold, or by requiring that project portfolios demonstrate measurable reductions in accessibility inequality over time. More broadly, because TI-acs can be recomputed as the network expands, it can support outcome-based accountability: agencies can track whether public investments are closing gaps in behaviorally realized access rather than simply increasing the count of installed ports.

Third, TI-acs supports \emph{grid-aware} charging policy by explicitly resolving accessibility across time-of-day. By decomposing accessibility into peak, off-peak, and super off-peak windows, TI-acs can help identify where and when time-varying price signals or managed-charging programs are most likely to succeed without imposing excessive burden on drivers. 
Finally, TI-acs suggests a pathway toward more transparent, participatory, and privacy-conscious planning. A practical policy workflow is to treat TI-acs as a \emph{decision-support layer} that complements, rather than replaces, standard siting considerations: land use, community preferences, safety, reliability, and long-run operations and maintenance. As charging networks and travel behavior evolve, agencies can update TI-acs priority maps periodically (e.g., annually) to ensure that deployment strategies remain aligned with changing mobility patterns and with the objective of equitable and effective access.

\section{Conclusions}

In this study, we propose a novel trajectory-integrated accessibility metric (TI-acs) to provide a comprehensive assessment of public EV charging accessibility.
Our empirical analysis demonstrates a steady improvement in public charger accessibility in the San Francisco Bay Area over the past decade. Measured in terms of accessible hours (the time individuals spend within 1 km of a public charger), Bay Area residents currently experience an average of 7.5 hours for L2 chargers and 5.2 hours for DCFC per day. Part of these periods coincide with grid operation hours with high renewable penetration, highlighting the potential for integrating demand-responsive charging programs. However, significant spatial disparities and racial inequities persist. Hispanic-majority communities exhibit substantially lower accessibility compared to White-majority communities, both in total and across different types of activity locations.

Methodologically, this work introduces novel perspectives on accessibility analysis. At the microscopic scale, by integrating individuals' daily stays and movements across various locations and time-of-use periods, our metric captures charging opportunities embedded in everyday activities. This granularity enables disaggregation by location types (home, workplace, and other destinations) and alignment with peak and off-peak grid hours. At the macroscopic scale, we evaluate how accessibility varies across communities and how it has evolved over time, providing insights into the effectiveness and equity of infrastructure deployment. This framework holds potential for application to other critical societal infrastructures and urban services beyond the context of EV charging.

By integrating detailed mobility patterns with public charging infrastructure data, our framework offers a more nuanced and actionable understanding of public EV charging accessibility. The findings emphasize the urgent need for equitable infrastructure expansion to ensure a more inclusive and sustainable transition to electric mobility.

\section*{Data \& Code Availability}
Aggregate data and analysis code needed to reproduce the results in this paper are available at: \url{https://github.com/eCal-UCB/TI-acs-public}. Due to copyright and privacy restrictions, the raw trajectory data cannot be publicly released.

\section*{Acknowledgement}
The authors thank \emph{the Salata Institute} at Harvard University and MIT \emph{CEEPR} for organizing \emph{workshop on the Economics and Policy of Electric Transportation Charging Infrastructure},
where Yi Ju was invited to present the research and received many constructive feedback.
The authors also thank Mr. \emph{Bhuvan Atluri}, Dr. \emph{Yunhan Zheng},  and Dr. \emph{Xinyi Wang} for their insightful comments on early versions of the manuscript.
High performance computing resources were provided by UC Berkeley BRC computing service (aka Savio).

\section*{Declaration of LLM usage in writing}
The authors used LLMs (primarily GPT-4o and GPT 5.2 Thinking) for enhancing the readability of the manuscript. The authors reviewed and edited the content as needed and take full responsibility for the content of the publication.

\printbibliography[title=References]

\newrefsection
\newpage

\section*{Appendix A: Supplementary plots}\label{sec:appex_A}

\section*{Appendix B: Preprocessing trajectories simulated by TimeGeo}\label{sec:appex_B}
The mobility profile generated directly from original TimeGeo simulator consists of a list of stays represented as \verb|(time, type, lon, lat)| for each individual. Here, \verb|time| is an integer in the set ${1, ..., 1008}$, indicating the time slot index for 10-minute intervals throughout the week. The \verb|type| is one of ``home", ``work", or ``other", and \verb|lon|, \verb|lat| represent the longitude and latitude of each stay location.

The original TimeGeo data does not explicitly model travel time. Instead, it assumes instant location transition as ``flights'' that do not consume time.
To estimate more realistic stay durations, we consider the start of the time slots to be the moment when the individual departs from their previous stay and moves to the next. Therefore, the exact start (arrival) time of a stay is the start of its associated time slot \emph{plus the travel time} from the last stay (except for the first stay, which starts at the 0th min).

To obtain the travel time between consecutive stays, we use the \verb|OSMnx| and \verb|networkx| packages on a ``drive'' map of the Bay Area, downloaded from \verb|OpenStreetMap|. This process involves matching stay locations to the nearest nodes on the graph and computing the shortest path between them using Dijkstra's algorithm, weighted by travel time. We add a constant time buffer of 6 minutes to the travel time given by shortest path routing to represent the required time of starting, stopping, and parking the vehicle. The routing calculations are computationally intensive, so we employ several algorithmic optimizations to improve efficiency on parallel computing systems.
For trips with no paths found on the road network (approximately 0.4\% of total trips), we approximate the trip distances and travel time based on their great circle distances and overall statistics of those matched trips.

Since TimeGeo does not explicitly account for travel time as a mobility constraint, approximately 3\% of all stays have an actual start time that is later than their end time due to long travel durations from the previous location. 
To address this inconsistency, for each stay whose duration is less than 5 minutes or even negative, we either advance its arrival time by 5 minutes (and adjust the departure time of the preceding stay accordingly to accommodate the required travel time) if the previous stay is long enough, or delay its departure time by 5 minutes (and adjust the arrival time of the subsequent stay accordingly) if the next stay allows for the adjustment.
We repeat this process through several iterations, successfully resolving more than 99\% of the inconsistencies. For the remaining cases, we simply shorten the travel time to ensure that every stay has a minimum duration of at least 5 minutes.

After postprocessing the TimeGeo data as described above, the time information of each stay can be compactly defined by \verb|start,end| as its start and end time in a week (round in minutes, in 0, ..., 10080).

\section*{Appendix C: Demographic variables and their codes from ACS}\label{sec:appex_C}
\begin{table}[h!]
\centering
\caption{Demographic characteristics used in the study.}
\label{tab:acs}
\normalsize
\resizebox{0.8\textwidth}{!}{%
\begin{tabular}{@{}ll | l@{}}
\toprule
\textbf{table} & \textbf{2022 code} & \textbf{variable description}                                       \\ \midrule
DP05           & 0001E              & Total population                                                    \\
DP05           & 0021E              & Total population above 18 years old (inclusive)                     \\
DP05           & 0073PE             & Percentage of Hispanic (or Latino) population (of any race)         \\
DP05           & 0079PE             & Percentage of White population (not Hispanic)                       \\
DP05           & 0080PE             & Percentage of Black (or African American) population (not Hispanic) \\
DP05           & 0082PE             & Percentage of Asian population (not Hispanic)                       \\
S1903          & C03\_001E          & Household median income                                             \\
S1101          & C01\_015E          & Percentage of households living in 1-unit structures                \\
S1101          & C01\_016E          & Percentage of households living in 2 or more unit structures        \\
S1101          & C01\_017E          & Percentage of households living in mobile homes                     \\
S1101          & C01\_018E          & Percentage of households who are owners of their housing units      \\
S1101          & C01\_019E          & Percentage of households who are renters of their housing units     \\ \bottomrule
\end{tabular}%
}
\end{table}

We include the variable code used in ACS 2022 in Table.~\ref{tab:acs}. Please note that for the same variable, its code may be different across different years.
\printbibliography[title=Supplementary References]

@article{zheng2024effects,
  title={Effects of electric vehicle charging stations on the economic vitality of local businesses},
  author={Zheng, Yunhan and Keith, David R and Wang, Shenhao and Diao, Mi and Zhao, Jinhua},
  journal={Nature Communications},
  volume={15},
  number={1},
  pages={7437},
  year={2024},
  publisher={Nature Publishing Group UK London}
}

@article{liang2023effects,
  title={Effects of expanding electric vehicle charging stations in California on the housing market},
  author={Liang, Jing and Qiu, Yueming and Liu, Pengfei and He, Pan and Mauzerall, Denise L},
  journal={Nature Sustainability},
  volume={6},
  number={5},
  pages={549--558},
  year={2023},
  publisher={Nature Publishing Group UK London}
}

@article{hsu2021public,
  title={Public electric vehicle charger access disparities across race and income in California},
  author={Hsu, Chih-Wei and Fingerman, Kevin},
  journal={Transport Policy},
  volume={100},
  pages={59--67},
  year={2021},
  publisher={Elsevier}
}

@article{powell2022charging,
  title={Charging infrastructure access and operation to reduce the grid impacts of deep electric vehicle adoption},
  author={Powell, Siobhan and Cezar, Gustavo Vianna and Min, Liang and Azevedo, In{\^e}s ML and Rajagopal, Ram},
  journal={Nature Energy},
  volume={7},
  number={10},
  pages={932--945},
  year={2022},
  publisher={Nature Publishing Group}
}

@article{xu2018planning,
  title={Planning for electric vehicle needs by coupling charging profiles with urban mobility},
  author={Xu, Yanyan and {\c{C}}olak, Serdar and Kara, Emre C and Moura, Scott J and Gonz{\'a}lez, Marta C},
  journal={Nature Energy},
  volume={3},
  number={6},
  pages={484--493},
  year={2018},
  publisher={Nature Publishing Group UK London}
}

@article{wu2024planning,
  title={Planning charging stations for 2050 to support flexible electric vehicle demand considering individual mobility patterns},
  author={Wu, Jiaman and Powell, Siobhan and Xu, Yanyan and Rajagopal, Ram and Gonzalez, Marta C},
  journal={Cell Reports Sustainability},
  volume={1},
  number={1},
  year={2024},
  publisher={Elsevier}
}

@article{jiang2016timegeo,
  title={The TimeGeo modeling framework for urban mobility without travel surveys},
  author={Jiang, Shan and Yang, Yingxiang and Gupta, Siddharth and Veneziano, Daniele and Athavale, Shounak and Gonz{\'a}lez, Marta C},
  journal={Proceedings of the National Academy of Sciences},
  volume={113},
  number={37},
  pages={E5370--E5378},
  year={2016},
  publisher={National Acad Sciences}
}

@article{boeing2024modeling,
  title={Modeling and Analyzing Urban Networks and Amenities with OSMnx},
  author={Boeing, Geoff},
  year={2024}
}

@techreport{hagberg2008exploring,
  title={Exploring network structure, dynamics, and function using NetworkX},
  author={Hagberg, Aric and Swart, Pieter J and Schult, Daniel A},
  year={2008},
  institution={Los Alamos National Laboratory (LANL), Los Alamos, NM (United States)}
}

@misc{us_department_of_energy_2024,
  author       = {{U.S. Department of Energy's Alternative Fuels Data Center}},
  title        = {{Electric Vehicle Charging Station Locations}},
  howpublished = {\url{https://afdc.energy.gov/fuels/electricity_locations.html}},
  note         = {Accessed: June 2024},
}

@misc{cec_ev_chargers_2024,
  author       = {{California Center for Sustainable Energy}},
  title        = {California Air Resources Board clean vehicle rebate project},
  howpublished = {\url{https://cleanvehiclerebate.org/en/rebate-map}},
  note         = {Accessed: June 2024},
}

@misc{ACCII_2024,
  author       = {{California Air Resources Board}},
  title        = {{Advanced Clean Cars II}},
  howpublished = {\url{https://ww2.arb.ca.gov/our-work/programs/advanced-clean-cars-program/advanced-clean-cars-ii}},
  note         = {Accessed: September 2024},
}

@misc{cec_ev_sales_2024,
  author       = {{California Energy Commission}},
  title        = {{New ZEV Sales in California}},
  howpublished = {\url{https://www.energy.ca.gov/data-reports/energy-almanac/zero-emission-vehicle-and-infrastructure-statistics-collection/new-zev}},
  note         = {Accessed: September 2024},
}

@misc{cvrp_2024,
  author       = {{California Energy Commission}},
  title        = {{Electric Vehicle Chargers in California}},
  howpublished = {\url{https://www.energy.ca.gov/data-reports/energy-almanac/zero-emission-vehicle-and-infrastructure-statistics-collection/electric}},
  note         = {Accessed: September 2024},
}

@misc{us_acs_2024,
  author       = {{United States Census Bureau}},
  title        = {{American Community Survey Data}},
  howpublished = {\url{https://data.census.gov/}},
  note         = {Accessed: September 2024},
}

@misc{ca_geoportal_2024,
  author       = {{California State Geoportal}},
  title        = {{California Zip Codes}},
  howpublished = {\url{https://gis.data.ca.gov/datasets/CDEGIS::california-zip-codes}},
  note         = {Accessed: June 2024},
}

@inproceedings{seabold2010statsmodels,
  title={statsmodels: Econometric and statistical modeling with python},
  author={Seabold, Skipper and Perktold, Josef},
  booktitle={9th Python in Science Conference},
  year={2010},
}

@article{bauer2021charging,
  title={Charging up America: Assessing the growing need for US charging infrastructure through 2030},
  author={Bauer, Gordon and Hsu, Chih-Wei and Nicholas, Mike and Lutsey, Nic},
  journal={White Paper ICCT},
  year={2021}
}

@inproceedings{bibra2022electric,
  title={Electric vehicle outlook 2022},
  author={Bibra, E and Connelly, E and Dhir, S and Drtil, M and Henriot, P and Hwang, I and Le Marois, J and McBain, S and Paoli, L and Teter, J},
  year={2022},
  organization={IEA}
}

@article{zhang2024sustainable,
  title={Sustainable plug-in electric vehicle integration into power systems},
  author={Zhang, Hongcai and Hu, Xiaosong and Hu, Zechun and Moura, Scott J},
  journal={Nature Reviews Electrical Engineering},
  volume={1},
  number={1},
  pages={35--52},
  year={2024},
  publisher={Nature Publishing Group UK London}
}

@article{elmallah2022can,
  title={Can distribution grid infrastructure accommodate residential electrification and electric vehicle adoption in Northern California?},
  author={Elmallah, Salma and Brockway, Anna M and Callaway, Duncan},
  journal={Environmental Research: Infrastructure and Sustainability},
  volume={2},
  number={4},
  pages={045005},
  year={2022},
  publisher={IOP Publishing}
}

@article{navidi2023coordinating,
  title={Coordinating distributed energy resources for reliability can significantly reduce future distribution grid upgrades and peak load},
  author={Navidi, Thomas and El Gamal, Abbas and Rajagopal, Ram},
  journal={Joule},
  volume={7},
  number={8},
  pages={1769--1792},
  year={2023},
  publisher={Elsevier}
}

@article{brockway2021inequitable,
  title={Inequitable access to distributed energy resources due to grid infrastructure limits in California},
  author={Brockway, Anna M and Conde, Jennifer and Callaway, Duncan},
  journal={Nature Energy},
  volume={6},
  number={9},
  pages={892--903},
  year={2021},
  publisher={Nature Publishing Group UK London}
}

@incollection{uscb1994census,
  author       = "U.S. Census Bureau",
  title        = "Census Tracts and Block Numbering Areas",
  booktitle    = "Geographic Areas Reference Manual",
  year         = 1994,
  publisher    = "U.S. Census Bureau",
  chapter      = 10
}

@misc{uscb2020tracts,
  author       = "{U.S. Census Bureau}",
  title        = "{2020 TIGER/Line Shapefile: Census Tracts}",
  year         = 2020,
  url          = "https://www.census.gov/cgi-bin/geo/shapefiles/index.php?year=2020&layergroup=Census+Tracts",
  note         = "Accessed: June 2024"
}

@article{lou2024income,
  title={Income and racial disparity in household publicly available electric vehicle infrastructure accessibility},
  author={Lou, Jiehong and Shen, Xingchi and Niemeier, Deb A and Hultman, Nathan},
  journal={Nature Communications},
  volume={15},
  number={1},
  pages={5106},
  year={2024},
  publisher={Nature Publishing Group UK London}
}

@article{falchetta2021electric,
  title={Electric vehicle charging network in Europe: An accessibility and deployment trends analysis},
  author={Falchetta, Giacomo and Noussan, Michel},
  journal={Transportation Research Part D: Transport and Environment},
  volume={94},
  pages={102813},
  year={2021},
  publisher={Elsevier}
}

@article{carlton2024electric,
  title={Electric vehicle charging equity and accessibility: A comprehensive United States policy analysis},
  author={Carlton, Gregory J and Sultana, Selima},
  journal={Transportation Research Part D: Transport and Environment},
  volume={129},
  pages={104123},
  year={2024},
  publisher={Elsevier}
}

@article{khan2022inequitable,
  title={Inequitable access to EV charging infrastructure},
  author={Khan, Hafiz Anwar Ullah and Price, Sara and Avraam, Charalampos and Dvorkin, Yury},
  journal={The Electricity Journal},
  volume={35},
  number={3},
  pages={107096},
  year={2022},
  publisher={Elsevier}
}

@article{jiao2024toward,
  title={Toward an equitable transportation electrification plan: Measuring public electric vehicle charging station access disparities in Austin, Texas},
  author={Jiao, Junfeng and Choi, Seung Jun and Nguyen, Chris},
  journal={Plos one},
  volume={19},
  number={9},
  pages={e0309302},
  year={2024},
  publisher={Public Library of Science San Francisco, CA USA}
}

@article{gazmeh2024understanding,
  title={Understanding the opportunity-centric accessibility for public charging infrastructure},
  author={Gazmeh, Hossein and Guo, Yuntao and Qian, Xinwu},
  journal={Transportation Research Part D: Transport and Environment},
  volume={131},
  pages={104222},
  year={2024},
  publisher={Elsevier}
}

@article{kontou2019understanding,
  title={Understanding the linkage between electric vehicle charging network coverage and charging opportunity using GPS travel data},
  author={Kontou, Eleftheria and Liu, Changzheng and Xie, Fei and Wu, Xing and Lin, Zhenhong},
  journal={Transportation Research Part C: Emerging Technologies},
  volume={98},
  pages={1--13},
  year={2019},
  publisher={Elsevier}
}

@article{esmaili2024assessing,
  title={Assessing the spatial distributions of public electric vehicle charging stations with emphasis on equity considerations in King County, Washington},
  author={Esmaili, Arsalan and Oshanreh, Mohammad Mehdi and Naderian, Shakiba and MacKenzie, Don and Chen, Cynthia},
  journal={Sustainable Cities and Society},
  volume={107},
  pages={105409},
  year={2024},
  publisher={Elsevier}
}

@article{peng2024analytical,
  title={An analytical framework for assessing equitable access to public electric vehicle chargers},
  author={Peng, Zhenhan and Wang, Matthew Wan Hong and Yang, Xiong and Chen, Anthony and Zhuge, Chengxiang},
  journal={Transportation Research Part D: Transport and Environment},
  volume={126},
  pages={103990},
  year={2024},
  publisher={Elsevier}
}

@article{roy2022examining,
  title={Examining spatial disparities in electric vehicle charging station placements using machine learning},
  author={Roy, Avipsa and Law, Mankin},
  journal={Sustainable cities and society},
  volume={83},
  pages={103978},
  year={2022},
  publisher={Elsevier}
}

@article{zhang2025equity,
  title={Equity evaluation of community-based public EV charging services--A case study of the Sacramento region},
  author={Zhang, Yunteng and Fan, Yueyue},
  journal={Energy Policy},
  volume={198},
  pages={114495},
  year={2025},
  publisher={Elsevier}
}

@misc{qian2025accessibility,
  author       = {Xinwu Qian and Hossein Gazmeh and Mario L. Small and Qi Wang and Yuntao Guo},
  title        = {The Accessibility and Inaccessibility of Urban Public Charging Stations},
  howpublished = {Preprint, Research Square},
  note         = {Version 1, available at \url{https://doi.org/10.21203/rs.3.rs-5702771/v1}},
  year         = {2025},
  month        = {January},
  day          = {24},
  doi          = {10.21203/rs.3.rs-5702771/v1}
}

@article{testi2024big,
  title={Big mobility data reveals hyperlocal air pollution exposure disparities in the Bronx, New York},
  author={Testi, Iacopo and Wang, An and Paul, Sanjana and Mora, Simone and Walker, Erica and Nyhan, Marguerite and Duarte, F{\'a}bio and Santi, Paolo and Ratti, Carlo},
  journal={Nature Cities},
  volume={1},
  number={8},
  pages={512--521},
  year={2024},
  publisher={Nature Publishing Group US New York}
}

@article{wang2021access,
  title={Access to urban parks: Comparing spatial accessibility measures using three GIS-based approaches},
  author={Wang, Siqin and Wang, Mingshu and Liu, Yan},
  journal={Computers, environment and urban systems},
  volume={90},
  pages={101713},
  year={2021},
  publisher={Elsevier}
}

@article{xu2025using,
  title={Using human mobility data to quantify experienced urban inequalities},
  author={Xu, Fengli and Wang, Qi and Moro, Esteban and Chen, Lin and Salazar Miranda, Arianna and Gonz{\'a}lez, Marta C and Tizzoni, Michele and Song, Chaoming and Ratti, Carlo and Bettencourt, Luis and others},
  journal={Nature Human Behaviour},
  pages={1--11},
  year={2025},
  publisher={Nature Publishing Group UK London}
}

@article{yabe2024enhancing,
  title={Enhancing human mobility research with open and standardized datasets},
  author={Yabe, Takahiro and Luca, Massimiliano and Tsubouchi, Kota and Lepri, Bruno and Gonzalez, Marta C and Moro, Esteban},
  journal={Nature Computational Science},
  volume={4},
  number={7},
  pages={469--472},
  year={2024},
  publisher={Nature Publishing Group US New York}
}

@article{yu2023california,
  title={California’s zero-emission vehicle adoption brings air quality benefits yet equity gaps persist},
  author={Yu, Qiao and He, Brian Yueshuai and Ma, Jiaqi and Zhu, Yifang},
  journal={Nature Communications},
  volume={14},
  number={1},
  pages={7798},
  year={2023},
  publisher={Nature Publishing Group UK London}
}

@article{ledna2022support,
  title={How to support EV adoption: Tradeoffs between charging infrastructure investments and vehicle subsidies in California},
  author={Ledna, Catherine and Muratori, Matteo and Brooker, Aaron and Wood, Eric and Greene, David},
  journal={Energy Policy},
  volume={165},
  pages={112931},
  year={2022},
  publisher={Elsevier}
}

@misc{iea2024evoutlook,
  author       = {{International Energy Agency}},
  title        = {Global EV Outlook 2024},
  year         = {2024},
  howpublished = {\url{https://www.iea.org/reports/global-ev-outlook-2024}},
  note         = {IEA, Paris. Licence: CC BY 4.0},
  organization = {International Energy Agency}
}

@misc{plug2024survey,
  author       = {{Plug In America}},
  title        = {2024 EV Driver Annual Survey Report},
  year         = {2024},
  howpublished = {\url{https://pluginamerica.org/survey/2024-ev-driver-survey/}}
}

@article{lee2019buying,
  title={Who is buying electric vehicles in California? Characterising early adopter heterogeneity and forecasting market diffusion},
  author={Lee, Jae Hyun and Hardman, Scott J and Tal, Gil},
  journal={Energy Research \& Social Science},
  volume={55},
  pages={218--226},
  year={2019},
  publisher={Elsevier}
}

@article{bashar2026evaluating,
  title={Evaluating spatial disparities in public EV charging infrastructure across the United States},
  author={Bashar, TM Junaid and Tao, Ran and Fernandes, Caleb and Jiao, Zhenzhi},
  journal={Journal of Transport Geography},
  volume={130},
  pages={104507},
  year={2026},
  publisher={Elsevier}
}

@article{mehditabrizi2025route,
  title={En route and home proximity in EV charging accessibility: a spatial equity analysis},
  author={Mehditabrizi, Asal and Tahmasbi, Behnam and Namadi, Saeed Saleh and Cirillo, Cinzia},
  journal={Transportation Research Part D: Transport and Environment},
  volume={146},
  pages={104910},
  year={2025},
  publisher={Elsevier}
}

@article{wang2026leveraging,
  title={Leveraging commuting patterns and workplace charging to advance equitable EV charger access},
  author={Wang, Ruiting and Kwon, Ha-Kyung and Jordan, Katherine H and Moura, Scott J and Boloor, Madhur and Machala, Michael L},
  journal={Transportation Research Part D: Transport and Environment},
  volume={150},
  pages={105089},
  year={2026},
  publisher={Elsevier}
}

@article{cai2026revisiting,
  title={Revisiting electric vehicle charging station accessibility: A home and workplace dual-scenario perspective},
  author={Cai, Leshan and Cui, Qinyu and Yang, Ziqi and Liu, Jiayu and Zhang, Kaihan},
  journal={Journal of Transport Geography},
  volume={130},
  pages={104488},
  year={2026},
  publisher={Elsevier}
}

@article{farnood2026proximity,
  title={Proximity planning for urban electrification: Walkable access to EV charging infrastructure in Montreal},
  author={Farnood, Ahad and Khorramisarvestani, Sepideh and Cucuzzella, Carmela and Gopakumar, Govind and Eicker, Ursula},
  journal={Journal of Urban Mobility},
  volume={9},
  pages={100174},
  year={2026},
  publisher={Elsevier}
}

@article{ozturk2025mesoscopic,
  title={A mesoscopic model of vehicular emissions informed by direct measurements and mobility science},
  author={{\"O}zt{\"u}rk, Ay{\c{s}}e Tu{\u{g}}ba and Kasliwal, Aparimit and Fitzmaurice, Helen and Kavvada, Olga and Calvez, Philippe and Cohen, Ronald C and Gonz{\'a}lez, Marta C},
  journal={Sustainable Cities and Society},
  volume={129},
  pages={106421},
  year={2025},
  publisher={Elsevier}
}

@article{li2022spatial,
  title={Spatial equity analysis of urban public services for electric vehicle charging—Implications of Chinese cities},
  author={Li, Guijun and Luo, Tanxiaosi and Song, Yanqiu},
  journal={Sustainable Cities and Society},
  volume={76},
  pages={103519},
  year={2022},
  publisher={Elsevier}
}

@article{hamim2025real,
  title={Real-world charging data reveals spatial heterogeneity in electric vehicle charging station usage},
  author={Hamim, Omar Faruqe and Chen, Xiaowei and Ukkusuri, Satish V},
  journal={Sustainable Cities and Society},
  pages={107045},
  year={2025},
  publisher={Elsevier}
}

@article{hu2025growth,
  title={Growth patterns and factors of electric vehicle charging infrastructure for sustainable development},
  author={Hu, Hongyu and Zhao, Dong and Zockaie, Ali and Ghamami, Mehrnaz},
  journal={Sustainable Cities and Society},
  volume={126},
  pages={106417},
  year={2025},
  publisher={Elsevier}
}

@article{liu2025planning,
  title={Planning public electric vehicle charging stations to balance efficiency and equality: A case study in Wuhan, China},
  author={Liu, Chenxi and Liu, Lingbo and Peng, Zhenghong and Wu, Hao and Wang, Fahui and Jiao, Hongzan and Wang, Jingjing},
  journal={Sustainable Cities and Society},
  volume={124},
  pages={106314},
  year={2025},
  publisher={Elsevier}
}

% \appendices
% \input{content/88-appendix}
% \small
%  \bibliographystyle{IEEEtran}
%  \bibliography{content/references}

\end{document}